\documentclass[twocolumn]{aastex631}

\usepackage{enumitem}
\usepackage{multirow}
\usepackage{booktabs}
\usepackage{comment}
\usepackage{graphicx}
\usepackage{soul}
\usepackage{amsmath}
\usepackage{amssymb}
\usepackage{xspace}
\usepackage{xifthen}
\usepackage{eso-pic}
\graphicspath{{./figures/}{./}}
\usepackage{hyperref}
\setlist{nolistsep}

\shorttitle{The GD-1 Stream in \Gaia DR3}
\shortauthors{Tavangar and Price-Whelan}

\begin{document}

\definecolor{forestgreen}{HTML}{228B22}
\definecolor{urlblue}{HTML}{000000}

\newcommand{\ie}{i.e.,\xspace}
\newcommand{\eg}{e.g.,\xspace}
\newcommand{\etc}{etc.\xspace}
\newcommand{\etal}{et al.\xspace}
\newcommand{\vs}{vs.\xspace}
\newcommand{\super}[1]{\ensuremath{^{\textrm{#1}}}}
\newcommand{\sub}[1]{\ensuremath{_{\textrm{#1}}}}
\newcommand{\FIXME}[1]{{\bf \textcolor{red}{#1}}}
\newcommand{\CHECK}[1]{{\textcolor{orange}{#1}}}
\newcommand{\COMMENT}[3]{\textcolor{#1}{#2: #3}}
\newcommand{\NOTE}[1]{\textcolor{blue}{(#1)}}

\newcommand{\highlight}[1]{\sethlcolor{yellow}\hl{#1}}
\newcommand{\response}[1]{{}}
\newcommand{\IT}{{I\&T}\xspace}

\newcommand{\Gaia}{{\it Gaia}\xspace}
\newcommand{\SSSSS}{${S}^5$\xspace}

\newcommand{\NKNOWN}{\CHECK{four}\xspace}
\newcommand{\NSTREAMS}{\CHECK{eleven}\xspace}
\newcommand{\NCAND}{\CHECK{eleven}\xspace}
\newcommand{\NGLOB}{\CHECK{four}\xspace}

\newcommand{\sectionfootnote}[2]{%
  \renewcommand{\thefootnote}{\fnsymbol{footnote}}%
  \section[#1]{#1${}^\dagger$}%
  \footnotetext[2]{#2}%
  \setcounter{footnote}{0}%
  \renewcommand{\thefootnote}{\arabic{footnote}}%
}

\mathchardef\mhyphen="2D
\newcommand{\vect}[1]{\boldsymbol{#1}}
\newcommand{\roughly}{\ensuremath{ {\sim}\,} }
\newcommand{\gtr}{\ensuremath{ {>}\,} }
\newcommand{\less}{\ensuremath{ {<}\,} }
\newlength{\dhatheight}
\newcommand{\doublehat}[1]{%
    \settoheight{\dhatheight}{\ensuremath{\hat{#1}}}%
    \addtolength{\dhatheight}{-0.35ex}%
    \hat{\vphantom{\rule{1pt}{\dhatheight}}%
    \smash{\hat{#1}}}}
\newcommand{\code}[1]{\texttt{#1}\xspace}
\newcommand{\dd}{\ensuremath{\rm d}}

\newcommand{\unit}[1]{\ensuremath{\mathrm{\,#1}}\xspace}
\newcommand{\yr}{\unit{yr}}
\newcommand{\Gyr}{\unit{Gyr}}
\newcommand{\Myr}{\unit{Myr}}
\newcommand{\eV}{\unit{eV}}
\newcommand{\keV}{\unit{keV}}
\newcommand{\MeV}{\unit{MeV}}
\newcommand{\GeV}{\unit{GeV}}
\newcommand{\TeV}{\unit{TeV}}
\newcommand{\MB}{\unit{MB}}
\newcommand{\GB}{\unit{GB}}
\newcommand{\TB}{\unit{TB}}
\newcommand{\degree}{\ensuremath{{}^{\circ}}\xspace}
\newcommand{\degrees}{\degree}
\newcommand{\mas}{\unit{mas}}
\newcommand{\masyr}{\unit{mas}~\unit{yr}^{-1}\xspace}
\newcommand{\amin}{\unit{arcmin}}
\newcommand{\asec}{\unit{arcsec}}
\newcommand{\angstrom}{\unit{\AA}}
\newcommand{\um}{\unit{$\mu$m}}
\newcommand{\cm}{\unit{cm}}
\newcommand{\km}{\unit{km}}
\newcommand{\kms}{\km \second^{-1}}
\newcommand{\pc}{\unit{pc}}
\newcommand{\kpc}{\unit{kpc}}
\newcommand{\second}{\unit{s}}
\newcommand{\us}{\unit{$\mu$s}}
\newcommand{\photons}{\unit{ph}}
\newcommand{\photon}{\unit{ph}}
\newcommand{\sr}{\unit{sr}}
\newcommand{\Msolar}{\unit{M_\odot}}
\newcommand{\Msun}{\unit{M_\odot}}
\newcommand{\Mstar}{\unit{M_{*}}}
\newcommand{\Lsolar}{\unit{L_\odot}}
\newcommand{\Lsun}{\unit{L_\odot}}
\newcommand{\Lstar}{\unit{L_{*}}}
\newcommand{\Lum}{\ensuremath{ L }\xspace}
\newcommand{\Dsun}{\unit{D_\odot}}
\newcommand{\Rgc}{\ensuremath{R_{GC}}\xspace}
\newcommand{\cmcubes}{\ensuremath{\cm^{3}\second^{-1}}\xspace}
\newcommand{\magn}{\unit{mag}}
\newcommand{\mmag}{\unit{mmag}}
\newcommand{\e}{\unit{e^{-}}}
\newcommand{\rms}{\unit{rms}}
\newcommand{\pix}{\unit{pix}}
\newcommand{\rmspix}{\unit{rms/pix}}
\newcommand{\ermspix}{\e \rmspix}

\newcommand{\secref}[1]{Section~\ref{sec:#1}}
\newcommand{\appref}[1]{Appendix~\ref{app:#1}}
\newcommand{\tabref}[1]{Table~\ref{tab:#1}}
\newcommand{\tabrefs}[2]{Tables~\ref{tab:#1} and \ref{tab:#2}}
\newcommand{\figref}[1]{Figure~\ref{fig:#1}}
\newcommand{\figrefs}[2]{Figures~\ref{fig:#1} and \ref{fig:#2}}
\newcommand{\eqnref}[1]{Equation~\eqref{eqn:#1}}

\newcommand{\bandvar}[2][]{%
  \ifthenelse{\isempty{#1}}{\var{#2}}{\var{#2\_#1}}%
}
\newcommand{\spreadmodel}[1][]{\bandvar[#1]{spread\_model}}
\newcommand{\spreaderrmodel}[1][]{\bandvar[#1]{spreaderr\_model}}
\newcommand{\wavgspreadmodel}[1][]{\bandvar[#1]{wavg\_spread\_model}}
\newcommand{\classstar}[1][]{\bandvar[#1]{class\_star}}
\newcommand{\magauto}[1][]{\bandvar[#1]{mag\_auto}}
\newcommand{\magpsf}[1][]{\bandvar[#1]{mag\_psf}}
\newcommand{\magerrpsf}[1][]{\bandvar[#1]{magerr\_psf}}
\newcommand{\flags}[1][]{\bandvar[#1]{flags}}

\newcommand{\LCDM}{\ensuremath{\rm \Lambda CDM}\xspace}
\newcommand{\modulus}{\ensuremath{m - M}\xspace}
\newcommand{\mM}{\modulus}
\newcommand{\ra}{{\ensuremath{\alpha_{2000}}}\xspace}
\newcommand{\dec}{{\ensuremath{\delta_{2000}}}\xspace}
\newcommand{\age}{{\ensuremath{\tau}}\xspace}
\newcommand{\metal}{{\ensuremath{Z}}\xspace}
\newcommand{\major}{\ensuremath{a_h}\xspace}
\newcommand{\nobjs}{{sixteen}\xspace}
\newcommand{\feh}{{\ensuremath{\rm [Fe/H]}}\xspace}

\newcommand{\ngmix}{\code{ngmix}}
\newcommand{\SWARP}{\code{SWarp}}
\newcommand{\swarp}{\SWARP}
\newcommand{\SExtractor}{\code{SExtractor}}
\newcommand{\sextractor}{\SExtractor}
\newcommand{\PSFex}{\code{PSFex}}
\newcommand{\Astromatic}{\code{Astromatic}}
\newcommand{\HEALPix}{\code{HEALPix}}
\newcommand{\healpix}{\HEALPix}
\newcommand{\healpy}{\code{healpy}}
\newcommand{\DAOPHOT}{\code{DAOPHOT}}
\newcommand{\PARSEC}{\code{PARSEC}}
\newcommand{\mangle}{\code{mangle}}
\newcommand{\emcee}{\code{emcee}}
\newcommand{\ugali}{\code{ugali}}
\newcommand{\var}[1]{\ensuremath{\texttt{\MakeUppercase{#1}}}\xspace}
\newcommand{\nside}{\code{nside}}

\newcommand{\prior}{\ensuremath{\mathcal{P}}\xspace}
\newcommand{\Prob}{\ensuremath{\mathcal{P}}\xspace}
\newcommand{\ProbJ}{\ensuremath{\mathcal{P}(J)}\xspace}
\newcommand{\like}{\ensuremath{\mathcal{L}}\xspace} 
\newcommand{\plike}{\like_p\xspace}                    
\newcommand{\jlike}{\ensuremath{L}\xspace}          
\newcommand{\pjlike}{\jlike_p\xspace}         
\newcommand{\pseudolike}{ {\tilde{\like}} \xspace}   
\newcommand{\loglike}{\ensuremath{\log\like}\xspace}
\newcommand{\logpseudolike}{\ensuremath{\log\pseudolike}\xspace}
\newcommand{\lnlike}{\ensuremath{\ln\like}\xspace}
\newcommand{\lnpseudolike}{\ensuremath{\ln\pseudolike}\xspace}
\newcommand{\given}{\ensuremath{ \,|\, }\xspace}
\newcommand{\likefn}[2]{\ensuremath{ \like(#1 \given #2) }\xspace}
\newcommand{\plikefn}[2]{\ensuremath{ \plike(#1 \given #2) }\xspace}
\newcommand{\jlikefn}[2]{\ensuremath{ \jlike(#1 \given #2) }\xspace}
\newcommand{\data}{ \ensuremath{ \mathcal{D} }\xspace } 
\newcommand{\param}{\ensuremath{{\theta}}\xspace}
\newcommand{\params}{\ensuremath{\vect{\theta}}\xspace}
\newcommand{\sig}{\ensuremath{\mu}\xspace}
\newcommand{\bkg}{\ensuremath{\eta}\xspace}
\newcommand{\interest}{\ensuremath{\vect{\sig}}\xspace}
\newcommand{\nuisance}{\ensuremath{\vect{\bkg}}\xspace}
\newcommand{\signal}{\sig}
\newcommand{\pvalue}{\textit{p}-value\xspace}
\newcommand{\pdf}{PDF\xspace}
\newcommand{\pdfs}{PDFs\xspace}

\newcommand{\uspatial}{\ensuremath{u_s}}
\newcommand{\ucolor}{\ensuremath{u_c}}

\newcommand{\Jlike}{\ensuremath{\like_{J}}\xspace}
\newcommand{\Jsigma}{\ensuremath{\sigma_{i}}\xspace}
\newcommand{\Jtrue}{\Ji}
\newcommand{\Jobs}{\barJi}
\newcommand{\logtenJtrue}{\logtenJi}
\newcommand{\logtenJobs}{\barlogtenJi}

\newcommand{\Ji}{\ensuremath{J_i}\xspace}
\newcommand{\barJi}{\ensuremath{ {\overline{J_i}} }\xspace}
\newcommand{\logJi}{\ensuremath{{\log{(J_i)}}}\xspace}
\newcommand{\barlogJi}{\ensuremath{ {\overline{\log{(J_i)}}} }\xspace}
\newcommand{\logtenJi}{\ensuremath{{\log_{10}{(J_i)}}}\xspace}
\newcommand{\barlogtenJi}{\ensuremath{ {\overline{\log_{10}{(J_i)}}} }\xspace}

\newcommand{\ScienceTools}{\code{ScienceTools}}
\newcommand{\Sourcelike}{\code{Sourcelike}}
\newcommand{\gtlike}{\code{gtlike}}
\newcommand{\pointlike}{\code{pointlike}}
\newcommand{\Gtlike}{\code{Gtlike}}
\newcommand{\gtobssim}{\code{gtobssim}}
\newcommand{\DMFIT}{\code{DMFIT}}
\newcommand{\Pythia}{\code{Pythia}}
\newcommand{\GALPROP}{\code{GALPROP}}

\newcommand{\DM}{\ensuremath{\mathrm{DM}}}
\newcommand{\mDM}{\ensuremath{m_\DM}\xspace}
\newcommand{\mLSP}{\ensuremath{m_\LSP}\xspace}
\newcommand{\mChi}{\ensuremath{m_\chi}\xspace}

\newcommand{\sigmav}{\ensuremath{\langle \sigma v \rangle}\xspace}
\newcommand{\sigmavmax}{\ensuremath{\sigmav_{\max}}\xspace}
\newcommand{\sigmavT}{\ensuremath{\sigmav_{\rm T}}\xspace}
\newcommand{\tsigmav}{\ensuremath{\sigmav R^{2}}\xspace}
\newcommand{\LSP}{\ensuremath{\chi}\xspace}
\newcommand{\PhiPP}{\ensuremath{\Phi_{\rm PP}}\xspace}
\newcommand{\uubar}{\ensuremath{u \bar u}\xspace}
\newcommand{\ddbar}{\ensuremath{d \bar d}\xspace}
\newcommand{\ccbar}{\ensuremath{c \bar c}\xspace}
\newcommand{\ssbar}{\ensuremath{s \bar s}\xspace}
\newcommand{\bbbar}{\ensuremath{b \bar b}\xspace}
\newcommand{\ttbar}{\ensuremath{t \bar t}\xspace}
\newcommand{\ww}{\ensuremath{W^{+}W^{-}}\xspace}
\newcommand{\zz}{\ensuremath{Z^{0}Z^{0}}\xspace}
\newcommand{\gluglu}{\ensuremath{gg}\xspace}
\newcommand{\ee}{\ensuremath{e^{+}e^{-}}\xspace}
\newcommand{\mumu}{\ensuremath{\mu^{+}\mu^{-}}\xspace}
\newcommand{\tautau}{\ensuremath{\tau^{+}\tau^{-}}\xspace}
\newcommand{\relic}{\ensuremath{3\times10^{-26}\cm^{3}\second^{-1}}\xspace}

\newcommand{\Jfactor}{J-factor\xspace}
\newcommand{\Jfactors}{J-factors\xspace}
\newcommand{\JFactor}{J-Factor\xspace}
\newcommand{\Rmax}{\ensuremath{R_{V_{\rm max}}}\xspace}
\newcommand{\Vmax}{\ensuremath{V_{\rm max}}\xspace}
\newcommand{\rs}{\ensuremath{ r_{\rm s} }\xspace}
\newcommand{\rhos}{\ensuremath{ \rho_0 }\xspace}
\newcommand{\alphahalf}{\ensuremath{ \alpha_{\rm h} }\xspace}
\newcommand{\ahalf}{ \alphahalf }
\newcommand{\rhalf}{\ensuremath{ r_{\rm h} }\xspace}
\newcommand{\Mhalf}{\ensuremath{ M_{\rm h} }\xspace}
\newcommand{\Mtidal}{\ensuremath{M_{\rm tidal}}\xspace}
\newcommand{\alphas}{\ensuremath{ \alpha_{\rm s} }\xspace}
\newcommand{\boost}{\ensuremath{\mathcal{B}}\xspace}
\newcommand{\vdisp}{\ensuremath{\sigma_\star}\xspace}

\providecommand\physrep{\ref@jnl{Phys.~Rep.}}%
\providecommand\apjs{\ref@jnl{ApJS}}%
\providecommand{\jcap}{\ref@jnl{JCAP}}%

\newcommand{\true}[1]{\tilde{#1}}
\newcommand{\bs}[1]{\boldsymbol{#1}}
\newcommand{\spline}[2]{\ensuremath{\mathcal{S}^{(#1)}({#2})}}
\newcommand{\norm}[3]{\ensuremath{\mathcal{N}(#1 \given #2, #3)}}
\newcommand{\truncnorm}[4]{\ensuremath{\norm{#1}{#2}{#3}_{[#4]}}}
\newcommand{\uniform}[2]{\ensuremath{\mathcal{U}(#1 \given #2)}}

\title{
    Inferring the density and membership of stellar streams with flexible models: \\
    The GD-1 stream in \Gaia Data Release 3
}

\author[0000-0001-6584-6144]{Kiyan Tavangar}
\affiliation{Department of Astronomy, Columbia University, New York, NY 10027, USA}

\author[0000-0003-0872-7098]{Adrian~M.~Price-Whelan}
\affiliation{Center for Computational Astrophysics, Flatiron Institute, Simons Foundation, 162 Fifth Avenue, New York, NY 10010, USA}
\correspondingauthor{Kiyan Tavangar}
\email{k.tavangar@columbia.edu}

\begin{abstract}
Stellar streams provide one of the most promising avenues for constraining the global mass distribution of the Milky Way and the nature of dark matter (DM).
The stream stars' kinematic ``track'' enables inference of large-scale properties of the DM distribution, while density variations and anomalies provide information about local DM clumps (\eg from DM subhalos).
A full accounting of the density tracks and substructures within all $>100$ Milky Way stellar streams will therefore enable powerful new constraints on DM.
Here, we present a new, flexible framework for modeling stellar stream density and membership.
With it, one can empirically model a given stream in a variety of coordinate spaces (\eg on-sky position and velocity) using probability distributions, thereby generating membership probabilities. 
The most significant improvement over previous methods is the inclusion of off-track or non-Gaussian components to the stream density, meaning we can capture anomalous features (such as the GD-1 steam's spur).
We test our model on GD-1, where we characterize previously-known features and provide the largest catalog of probable member stars to date (1689 stars).
We then use the derived model to provide measurements of GD-1's density and kinematic tracks, velocity dispersion, as well as its initial and current mass.
Our framework (built on \texttt{JAX} and \texttt{numpyro}) provides a path toward uniform analysis of all Milky Way streams, enabling tight constraints on the Galactic mass distribution and its dark matter.
\end{abstract}

\section{Introduction} \label{sec:intro}

In the $\Lambda$-Cold Dark Matter ($\Lambda$CDM) cosmological paradigm, galaxies form through hierarchical structure formation.
Small matter overdensities in the early universe combine to form dark matter halos, which subsequently merge to create larger structures capable of hosting galaxies \citep[\eg][]{Peebles:1965, Press_Schechter:1974, White_Rees:1978, Blumenthal:1984, Davis:1985}.
There is strong evidence supporting this theory across a range of mass and length scales in the universe, from large-scale structure \citep[\eg][]{White_Frenk:1991, Springel:2005, Eisenstein:2005, Klypin:2011} to the nearby universe \citep[\eg][]{Bullock:2000, Drlica-Wagner:2015}.
For example, in our galaxy, observations indicate that over its history, the Milky Way has accreted dozens of dwarf galaxies and potentially hundreds of smaller objects such as globular clusters (GCs) \citep{Searle_Zinn:1978, Vasiliev:2019}.

As these objects fall into the Milky Way, the Galaxy's gravitational force acts on their constituent stars, sometimes unbinding them from their progenitors \citep{Toomre_Toomre:1972}.
This process, known as tidal stripping, forms structures known as stellar streams and shells (see \citealt{Bonaca:25} for a recent overview of streams in the Milky Way).
Observations of these shells and streams in the Galaxy over the past half-century constitute some of the strongest evidence for the hierarchical build-up of stellar halos and of \LCDM \citep[\eg][]{Odenkirchen:2001, Newberg:2002, Belokurov:06, Shipp:18, Malhan:18, Ibata:24}.

In the Milky Way alone, we now count nearly 150 stellar streams \citep{Mateu:2023}, with an explosion of stream discoveries over the past decade from surveys such as \Gaia \citep{Gaia:2016}, the Sloan Digital Sky Survey \citep{Kollmeier:2017}, and the Dark Energy Survey \citep{DES:2005}.
In the next decade, we expect telescopes such as the Vera Rubin Observatory and the Nancy Grace Roman Space Telescope to discover dozens more streams both in the Milky Way and external galaxies \citep{Pearson:22, Bonaca:25}.
Streams are useful objects for tackling two major research questions: understanding the nature of dark matter \citep{Johnston:02, Ibata:02, Siegal-Gaskins:08, Carlberg:09, Yoon:2011, Bonaca:2019, Drlica-Wagner:19, Foote:24, Hilmi:24} and constraining the shape of galaxy potentials \citep[\eg][]{Springel:99, Dubinski:99, Johnston:99, Johnston:05, Binney:08, Koposov:10, Law:10, Amorisco:15, Price-Whelan:2014, Sanders:2014, Erkal:2016b, Bovy:16, Shipp:21, Pearson:22a, Koposov:23, Ibata:24}.

For the former, streams allow us to constrain dark matter halos on smaller scales than most other objects in the Universe (i.e. dark matter subhalos with masses $M \gtrsim 10^6~\Msun$).
A stream on a circular orbit formed in a perfectly smooth potential will have a nearly homogeneous density distribution along its length and follow a smooth track in all astrometric coordinates, with density variations caused only by orbital pileups known as ``epicyclic over-densities'' \citep[\eg][]{Kupper:2012}.
Streams on more elliptical orbits have non-uniform mass loss along their orbits \citep[\eg][]{Fardal:2015, Penarrubia:08}.
This may cause inhomogeneities in the stream density, particularly close to the progenitor, but these inhomogeneities should occur predictably and regularly.
However, if a stream interacts with lumps in the potential (\eg small dark matter subhalos), these perturbations can create unpredictable density gaps, small-scale fluctuations in the astrometric stream tracks, or stream stars moving off the main track, all of which exist in observed streams.
Consequently, streams can be used to study dark matter subhalos below the threshold for galaxy formation $\lesssim 10^8 M_{\odot}$ \citep{Yoon:2011, Bonaca:14, Benitez-Llambay:20}.
Identifying subhalos in this mass range would
provide key observations of structure in a regime where \LCDM differs from competing theories of dark matter, such as fuzzy DM \citep{Hui:2017}, warm dark matter \citep[WDM;][]{Bond_Szalay:1983, Bode:2001}, and self-interacting DM \citep[SIDM;][]{Spergel_Steinhardt:2000, Elbert:2015}.
Streams therefore offer a straightforward conceptual path towards an improved understanding of dark matter.

Streams are also powerful tools with which to infer the global mass distribution of the Milky Way because they approximately trace orbits \citep{Bonaca_Hogg:2018}. 
Due to the increased precision from ensemble averaging stream star orbits, even a single well-measured stream can constrain the shape of its host galaxy's potential \citep{Law:10, Ibata:13, Vera-Ciro:13, Gibbons:14, Dierickx:17, Newberg:10, Price-Whelan:2014, Bowden:15, Kupper:15, Bovy:16, Pearson:22a} and those constraints only tighten when using multiple streams to derive potential parameters \citep[\eg][]{Erkal:2019, Shipp:2021, Koposov:23}.
Recently, \citet{Ibata:24} leveraged the combined power of 29 stellar streams to present a global mass model of the Milky Way and infer its shape.

Constraining the large-scale and small-scale properties of dark matter around the Milky Way requires excellent empirical models of observed streams.
This is a difficult problem because most streams have been orbiting their host potentials for billions of years, and they can differ from idealized simulations in a multitude of ways.
A number of streams around the Milky Way, for example, display complex morphological features:
ATLAS-Aliqa Uma has a discontinuity in its track \citep{Li:2021}; Jhelum appears to be non-Gaussian in its matter distribution across the stream, with both a broad and narrow component \citep{Shipp:2018, Bonaca:19a}; and
GD-1 contains a ``spur'', a stream feature lying off its main track \citep{Price-Whelan_Bonaca:2018}.
Empirical models of stream density and membership must be capable of and effective at capturing these anomalies since they are features of great interest for constraints on dark matter and host potentials.

Over the past decades, many methods have been developed to study the density structure of observed streams \citep[\eg][]{Odenkirchen:2001, Erkal:2017, Koposov:2019, Li:2021, Ferguson:2022, Tavangar:22, Patrick:2022, Starkman:2024}.
These models have generally been effective at recovering gaps and under-densities, but ``off-track,'' non-Gaussian, or discontinuous features of streams have been more difficult to incorporate (for simplicity, future references to ``off-track'' features refer to all three of these types of features, unless otherwise noted).
In particular, off-track features are usually modeled by assuming their existence \citep[\eg][]{Starkman:2024}.
Such a model relies on prior knowledge of an off-track feature, most likely based on visual inspection of stream members.
Therefore, it is incapable of detecting previously undiscovered or faint off-track structures.
In this paper, we present an empirical model of stream density that does accommodate and can recover new off-track components.
With the model, we can generate stream tracks in kinematic coordinates (3D positions and velocities) and density as well as calculate membership probabilities for stars in the region.

Along with a description of our model framework and an associated \texttt{Python} package for constructing generic stream models, we present an application to the GD-1 stream, discovered by \citet{Grillmair_Dionatos:2006}.
GD-1 is one of the most promising streams for obtaining constraints on the Milky Way potential and especially the nature of dark matter \citep{Lux:2013, Bonaca_Hogg:2018, Lu:25}.
It is a dynamically cold, thin, and metal-poor globular cluster stream spanning approximately $100\degr$ across the sky \citep{deBoer:2018, Price-Whelan_Bonaca:2018}.
GD-1's proximity ($\approx 8$ \kpc) and high surface brightness mean it is a prime candidate for detailed observations of any inhomogeneities.
Indeed, GD-1 contains a few prominent such features: many gaps along the stream along with an off-track ``spur'' and an off-track ``blob'' \citep{Price-Whelan_Bonaca:2018}.
These latter two features are particularly exciting because their observable properties (\eg angle away from the stream, length, and proper motion) allow an excellent opportunity to constrain the perturbation that caused them.
A better constraint on this perturbation in GD-1 (and specifically an improved estimate of the mass of the perturber) could lead to a better understanding of dark matter \citep{Bonaca:2019}.

We intend our method to be used in examinations of known streams.
While the model could in theory discover new streams, it is designed to function as a characterization tool for known streams, where we already have an idea of stream properties.
The paper is organized as follows.
In \secref{methods}, we detail the model.
In \secref{gd1_app}, we apply the model to GD-1 and show that we recover the best sample of stream stars to date.
We discuss the results in \secref{discussion} and finally conclude in \secref{conclusion}.

\section{Methods: A flexible density model for stellar streams} \label{sec:methods}

The goal of our modeling framework is to infer the intrinsic number density distribution or probability distribution function (\pdf) of stars in a stellar stream given noisy observations of its kinematic properties.
Here, we focus on the kinematics of stars, but note below that this will be extended to also model the stellar population (or color--magnitude) structure of a stream.
We consider observable kinematic coordinates like sky position $(\phi_1, \phi_2)$ (longitude and latitude in a spherical coordinate system rotated with respect to standard equatorial coordinates in the International Celestial Reference System), distance $d$ or distance modulus $\textrm{DM}$, proper motion components\footnote{In notation, we assume that the proper motion value in the longitude $\mu_{\phi_1}$ implicitly contains a multiplicative $\cos{\phi_2}$ factor.} $(\mu_{\phi_1}, \mu_{\phi_2})$, and radial (line-of-sight) velocity $v_r$.
For most streams, the rotated coordinate system we work in --- $(\phi_1, \phi_2)$ --- is chosen such that $\phi_1$ increases monotonically along the extent of the stream.
For brevity, we define $\mu_1 = \mu_{\phi_1}$ and $\mu_2 = \mu_{\phi_2}$.
In our notation\footnote{We summarize this notation in Table~\ref{table:notation}} below, we also define these (or potentially a specified subset of these coordinates) as the data vector $\bs{y} = (\phi_1, \phi_2, d, \mu_1, \mu_2, v_r)$.
Given noisy observations of these phase-space coordinates $\bs{y}$ for some sample of stars, we then want to infer the parameters $\bs{\theta}$ of a model for the probability density over the \emph{true} phase-space coordinates $\true{\bs{y}} = (\true{\phi}_1, \true{\phi}_2, \true{d}, \true{\mu}_1, \true{\mu}_2, \true{v}_r)$.

Of course, no practical data set will contain a pure sample of stars in a stellar stream.
We must therefore account for non-member ``background'' stars in our modeling setup.
We handle this by constructing a \emph{mixture model} for the probability density of observed phase-space coordinates.

Our flexible stream density model framework is publicly available as the Python package \texttt{stream-membership} on Github.\footnote{https://github.com/stellarstreams/stream-membership}
This package is built using \texttt{jax} \citep{jax:18} and \texttt{numpyro} \citep{Bingham:19, Phan:19}; \texttt{jax} enables automatic differentiation, just-in-time (JIT) compilation, and automatic vectorization of the model implementation, and \texttt{numpyro} provides a probabilistic model-building framework for implementing probability density models.
These libraries make our model fitting process efficient, user-friendly, and extensible, which we elaborate on in \secref{comparison}.

In the subsections below, we describe how we represent and parameterize the stream and background components of our density mixture model for our default stream model implementation.
However, we note that many of these details are customizable using any standard functionality from \texttt{jax} and \texttt{numpyro}.

\subsection{The Stream Model Component}

Many stellar streams around the Milky Way appear as over-densities of stars with spatial and velocity trends that vary smoothly along their length.
For our default stream density model, we therefore choose to factorize the joint \pdf over all phase-space coordinates into a term that specifies the linear density in stream longitude $\phi_1$, with all other phase-space coordinate distributions conditional on longitude $\phi_1$.
In detail, we factorize the joint \pdf for the stream component, $p_s(\true{\bs{y}} \given \bs{\theta}_s)$, as:
\begin{equation}
\begin{split}
p_s(\true{\bs{y}} \given \bs{\theta}_s) &=
    p_s(\true{\phi}_1 \given \bs{\theta}_s)
    \, p_s(\true{\phi}_2 \given \phi_1, \bs{\theta}_s)
    \, p_s(\true{d} \given \phi_1, \bs{\theta}_s) \\
    &\times
    p_s(\true{\mu}_1 \given \phi_1, \bs{\theta}_s)
    \, p_s(\true{\mu}_2 \given \phi_1, \bs{\theta}_s)
    \, p_s(\true{v}_r \given \phi_1, \bs{\theta}_s)
\end{split}
\end{equation}
where $\bs{\theta}_s$ are the parameters that specify the stream density model.
We have not yet specified the functional form of the coordinate distributions (e.g., $p_s(\true{\phi}_2 \given \phi_1, \bs{\theta}_s)$) or the parameters of the stream model component: We define these later when specifying our application to the GD-1 stream.

\subsection{Background Model Component} \label{sec:background}

To properly characterize the density of structures in the Milky Way halo, it is vital to characterize the density of background (or foreground) stars.
When modeling stellar membership of compact structures like globular clusters and dwarf galaxies, one can often assume a constant or linearly varying background density of stars because these structures cover a small area on the sky.
However, stellar streams are extended objects that can span tens or hundreds of degrees, meaning the background stellar density can vary substantially along a stream.
Our framework is flexible enough to allow for a variety of background models, from simple constant or linearly varying densities to more complex models that vary in multiple dimensions.
We represent the background component of our mixture model with the \pdf $p_b(\true{\bs{y}} \given \bs{\theta}_b)$, which has its own parameters $\bs{\theta}_b$.

\subsection{Off-track Model Component} \label{sec:off-track}

Many observed streams are now known to have complex morphological features such as gaps, spurs, blobs, breaks, and multi-component densities (see, e.g., \citealt{Bonaca:25}) that represent populations of stars that are co-moving with a stream but may not be well-represented by a single conditional density model as defined above.
We therefore generally add an additional component to our overall mixture model to represent these ``off-track'' stars.
We represent this mixture component with the \pdf $p_o(\true{\bs{y}} \given \bs{\theta}_o)$, which has its own parameters $\bs{\theta}_o$.
This component may have more complex joint dependencies, so we do not generically factorize to be fully conditional on $\phi_1$ as we did for the stream component.
However, we will often either explicitly tie some parameters of this component to the parameters of the stream component, or we will place strong priors on the parameters of this component to ensure that it does not interfere with the main stream component in regions where a 1D stream is well-defined.

\subsection{The Full Density Model} \label{sec:full}

In our default stream density model, we represent the full stellar density distribution as a mixture of the stream, background, and  off-track components so that the full \pdf for a star's true phase-space coordinates $\true{\bs{y}}$ is
\begin{equation} \label{eqn:total_prob_dens}
    p(\true{\bs{y}} \given \bs{\theta}) =
        \alpha_s \, p_s(\true{\bs{y}} \given \bs{\theta}_s) +
        \alpha_b \, p_b(\true{\bs{y}} \given \bs{\theta}_b) +
        \alpha_o \, p_o(\true{\bs{y}} \given \bs{\theta}_o)
\end{equation}
where $\bs{\theta} = (\bs{\theta}_s, \bs{\theta}_b, \bs{\theta}_o)$ and $\bs{\alpha} = (\alpha_s,\alpha_b, \alpha_o)$ are the mixing coefficients that specify the relative contributions of the stream, background, and off-track components to the overall density distribution (and $\alpha_s + \alpha_b + \alpha_o= 1$).
The likelihood of the observed data $\bs{y}$ given the density model parameters $\bs{\theta}$ is then
\begin{equation}
    p(\bs{y}, \true{\bs{y}} \given \bs{\theta}) =
        p(\bs{y} | \true{\bs{y}}) \,
            p(\true{\bs{y}} \given \bs{\theta})
\end{equation}
where in most cases we assume (as is standard) that the observational noise is Gaussian with a covariance matrix specified by the survey catalogs we use below (e.g., for \Gaia; \citealt{Hogg:2018}).
With this assumption,
\begin{equation}
    p(\bs{y} | \true{\bs{y}}) =
        \mathcal{N}(\bs{y} | \true{\bs{y}}, \Sigma_y)
\end{equation}
where $\mathcal{N}(\bs{y} | \true{\bs{y}}, \Sigma_y)$ represents the normal distribution
with mean $\true{\bs{y}}$ and the covariance matrix $\Sigma_y$ is specified by the
observational uncertainties in the data.
In practice, we have data and true phase-space coordinates for a sample of $N$ stars (indexed by $n$) so that the total likelihood $\mathcal{L}$ is
\begin{equation}
    \mathcal{L}(\{\true{\bs{y}}\}_N, \bs{\theta} \,;\,
        \{\bs{y}\}_N) =
        \prod_n^N p(\bs{y}_n | \true{\bs{y}}_n) \,
            p(\true{\bs{y}}_n \given \bs{\theta}) \label{eq:like} \quad .
\end{equation}

We are interested in inferring the parameters $\bs{\theta}$ of the density model and not
the true, per-star phase-space coordinates $\true{\bs{y}}_n$.
That is, we are most interested in the marginal posterior \pdf $p(\bs{\theta} \given
\{\bs{y}\}_N)$.
In the next section, we describe how we perform this inference using variational
inference.

\subsection{Inference} \label{sec:inference}

With the model scaffolding in place (details about our choices for parameterizing this model for GD-1 are shown in \secref{gd1_app}), our next goal is to infer the parameters $\bs{\theta}$ of the density model.
As we are interested in the stream density distribution and uncertainties in the parameters, we opt to simultaneously infer and marginalize over the true phase-space coordinates $\true{\bs{y}}$, the background density component parameters $\bs{\theta}_b$, and the mixture weight coefficients $\bs{\alpha}$.
That is, we are interested in the marginal posterior \pdf
\begin{equation}
\begin{split}
    p(\bs{\theta}_s, \bs{\theta}_o \given \{\bs{y}\}_N) =
        &\int \dd \bs{\theta}_b \, \dd \bs{\alpha} \, \dd^N \true{\bs{y}} \\
        &\times p(\bs{\theta}_s, \bs{\theta}_b, \bs{\theta}_o, \bs{\alpha}, \{\true{\bs{y}}\}_N \given \{\bs{y}\}_N)
\end{split}
\end{equation}
where the full posterior \pdf is specified by multiplying our likelihood (Equation~\ref{eq:like}) by prior \pdfs over all model parameters,
\begin{equation}
\begin{split}
    p(\bs{\theta}_s, &\bs{\theta}_b, \bs{\theta}_o, \bs{\alpha}, \{\true{\bs{y}}\}_N
        \given \{\bs{y}\}_N) \propto \\
        &\left[
            \prod_n^N p(\bs{y}_n | \true{\bs{y}}_n) \, p(\true{\bs{y}}_n
            \given \bs{\theta})\right] \,
        p(\bs{\theta}_s, \bs{\theta}_b, \bs{\theta}_o, \bs{\alpha}) \quad .
\end{split}
\end{equation}

A standard approach to this inference problem is to use a Markov Chain Monte Carlo (MCMC) method to sample the full posterior \pdf, and consider only values for the parameters of interest.
Here, however, our parameter space is high-dimensional (it scales with the number of stars we use in our inference) and MCMC would therefore be computationally prohibitive.
In special cases --- i.e. when the forms of the component distributions $p_s(\cdot)$, $p_b(\cdot)$, $p_o(\cdot)$ are Gaussian for all coordinates --- we can analytically marginalize over the true phase-space coordinates $\true{\bs{y}}$, greatly reducing the number of parameters we need to infer.
In general cases, this marginalization must be done numerically.
We therefore use variational inference to approximate the full posterior \pdf and enable this high-dimensional marginalization.

Variational inference \citep[VI;][]{Jordan:99, Wainwright:08, Blei:17} is a method for approximating the posterior \pdf by optimizing a simpler ``guide'' distribution that is easier to sample from.
VI works by maximizing the evidence lower bound (ELBO) of the guide distribution.
We use \texttt{numpyro} to perform the VI optimization and typically use a composition of independent normal distributions for each parameter as the guide distribution.
This choice is approximate in that it assumes that the posterior \pdf is Gaussian and cannot capture covariances between parameters in the posterior \pdf, but it is computationally efficient and often provides a conservative approximation to the true posterior \pdf.

\section{Application to the GD-1 Stream: Model Setup and Inference} \label{sec:gd1_app}

Having defined the general framework for our model, we now turn to a specific application to the GD-1 stellar stream.
GD-1 is a natural choice for this demonstration because of its length ($\approx 100\degree$), relative proximity (distance $d \approx 8 \kpc$), high surface brightness, and known density variations \citep[\eg][]{Ibata:24}.
It also contains off-track features (a spur and a blob; \citealt{Price-Whelan_Bonaca:2018}) that provide an excellent test for the major innovation of our method: modeling co-moving, off-track density components of streams.

We choose to model the stream in sky positions, proper motions, and radial velocity (\ie $\bs{y} = (\phi_1,\phi_2,\mu_1,\mu_2,v_r)$), but do not model the distance (or parallax) distribution along the stream.
For the model to function, every star must have measurements (including uncertainties) for each coordinate.
Every star has sky positions and proper motions but only a small fraction have radial velocities.
For the stars with no $v_r$ measurement, we assign $v_r = 0 \kms$ with a velocity uncertainty set to a very large number $\sigma_{v_r}=10^4 \kms$, which effectively treats these as missing measurements.
Additionally, the \Gaia errors on sky position are negligible, so we arbitrarily set those at $10^{-4}$ deg.

We do not model the stellar population (color--magnitude) of the stream in our probabilistic model.
We instead perform selections in color and magnitude to select metal-poor and distant stars.
This means we also do not model the stream's distance track: the stellar parallaxes are very low signal-to-noise and thus most distance information would come from modeling the color--magnitude distribution of the stream.

We release the code for this application to GD-1 in a separate repository\footnote{
https://github.com/ktavangar/gd1-dr3} from the more general \texttt{stream-membership} code.

\subsection{Data pre-selection} \label{sec:data}

For this analysis we use data from \Gaia's third data release \citep[DR3;][]{Gaia:2016,Gaia:2023} and PanSTARRS1's second data release \citep[PS1 DR2;][]{Chambers:2016}.
\Gaia is an all-sky stellar survey that provides positions and velocities for almost two billion stars in the Milky Way, allowing us to constrain their precise orbits.
PS1 is a photometric survey of the northern sky ($\delta > -30\degree$), which has a deeper magnitude limit than \Gaia and therefore has higher photometric precision.
We correct the PS1 photometry for extinction and then cross-match it to the \Gaia datasets. 
Henceforth, we use \Gaia astrometry and PS1 photometry unless otherwise noted.
We reduce the computational burden by filtering out stars that are highly unlikely to be part of the GD-1 stream with pre-selections on proper motions and the color--magnitude distribution of stars in this cross-matched dataset.
We generally use the \texttt{galstreams} track of GD-1 \citep{Mateu:2023} to help define these selections.

We consider the following two factors when forming the dataset on which the model is run.
First, a higher purity sample is helpful because it reduces the number of stars and allows easier separation of stream and background stars.
These improve the computational efficiency and model performance, respectively.
Second, a high completeness sample is necessary.
In particular, we cannot have cuts in any modeled coordinate $\bs{y}$ that excludes stream stars with specific properties $\bs{\hat{y}}$ because the stream component of our model will then be biased against including the area around $\bs{\hat{y}}$.
Given this, we make as many cuts to the input dataset as possible without a significant effect on stream completeness in the modeled coordinates.

We first select a spatial region around GD-1.
Previous works have identified GD-1 members in the $-90 \degree < \phi_1 < 10 \degree$ \citep[\eg][]{Price-Whelan_Bonaca:2018}.
To make sure we do not omit any stream members outside this $\phi_1$ range, we include an additional $10 \degree$ on either side in our data selection.
We then choose $-7.97\degree < \phi_2 < 3.35\degree$, which includes the entirety of the known stream and a few degrees above and below it, which are crucial for creating a background model.
We then make two initial cuts: a parallax cut of $\varpi < 1$ arcsecond to eliminate nearby stars and a magnitude cut of \Gaia $G < 20.5$.
The latter ensures that our density model does not depend on \Gaia's selection function since for $G < 20.5$ in this sky region, \Gaia is $\approx 100$\% complete with no spatial variation in the selection function \citep{Cantat-Gaudin:23}.

\begin{figure}[t!]
    \centering
    \includegraphics[width=\linewidth]{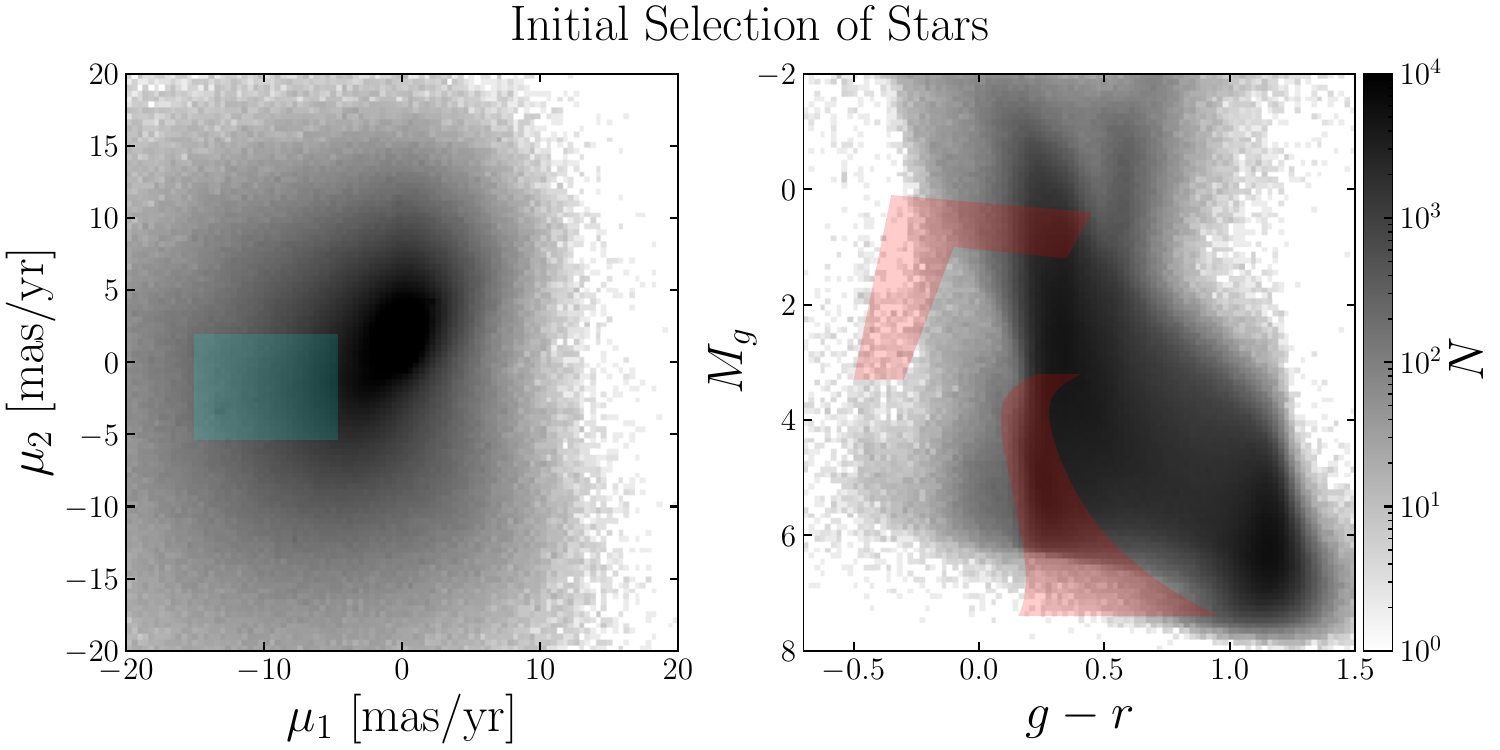}
\caption{
    The distribution of all stars in our \Gaia--PS1 cross-matched sample in the sky region near GD-1 (grayscale) and a visualization of our initial selections (blue and red polygons).
    The color scale of the grayscale images is logarithmic.
    \textbf{Left:} The distribution of proper motions in stream-aligned coordinates ($\phi_1, \phi_2$). The blue rectangle shows the initial selection we make in this space to create the input data for our model.
    \textbf{Right:} The distribution of stars in PS1 $g-r$ color and absolute $g$-band magnitude $M_g$, using a previously inferred distance track \citep{Valluri:25}.
    The red polygons show the selections we make to create the input data for our model.
}
\label{fig:gaia_cuts}
\end{figure}

We then make a generous proper motion cut to remove disk and general halo stars that are very unlikely to be part of GD-1.
We use results from \citet{Price-Whelan_Bonaca:2018} (henceforth PWB18), which published GD-1 tracks for $\mu_1$ and $\mu_2$ as a function of $\phi_1$ in the range $-90 \degree < \phi_1 < 10 \degree$.\footnote{These proper motion tracks, along with those for the other astrometric coordinates ($\phi_2, d, v_r$), are publicly available and easily accessible through the \texttt{galstreams} python package \citep{Mateu:2023}.}
Specifically, we extend the tracks to cover our full $\phi_1$ range by extrapolating from an interpolated spline of the PWB18 results.
We then take the minimum and maximum values of the extended $\mu_1$ and $\mu_2$ tracks and create a buffer of 2 mas/yr around those values.
This creates a rectangular selection in proper motion space shown in the left panel of \figref{gaia_cuts}.
We summarize our data cuts so far with:
\begin{itemize}
    \smallskip
    \item $-100 \degree < \phi_1 < 20 \degree$
    \smallskip
    \item $-7.97 \degree < \phi_2 < 3.35 \degree$
    \smallskip
    \item $\varpi<1$ arcsecond
    \smallskip
    \item $G < 20.5$ mag
    \smallskip
    \item $-15.10 \masyr < \mu_1 < -4.64 \masyr$
    \smallskip
    \item $-5.42 \masyr < \mu_2 < 1.93 \masyr$
    \smallskip
\end{itemize}

We now turn to our stellar population (color--magnitude) cut.
Since we do not model this space, we cannot bias the model by being incomplete in this space.
Therefore, we can make more stringent cuts and safely increase the purity to aid the stream and background separation as well as the computational efficiency, as mentioned above.
We assume GD-1's stars come from a single stellar population, which allows us to use a single isochrone as long as we account for the distance gradient along the stream.
To do so, we use the distance modulus track from \citet{Valluri:25}:
\begin{equation}
    \textrm{DM} = 0.0002440 \, \phi_1^2 + 0.02441 \, \phi_1 + 14.98
\end{equation}
This allows us to create a color--absolute magnitude diagram ($M_g$ \vs $g-r$), as shown in the right panel of \figref{gaia_cuts}.
We assume the stream lies along a Dotter isochrone \citep{Dotter:2016} with $\feh = -2.5$ dex and age $= 12$ \Gyr \citep{Valluri:25}.
This theoretical isochrone may not perfectly match the stream's photometric distribution, so we allow it to shift in both $M_g$ and $g-r$ (with its shape fixed) until it is the best fit to the stream's main sequence.
We then draw a polygon around the isochrone's main sequence ($M_g > 3.2$) to create our selection box.
The width of our selection region depends on the $M_g$ uncertainty as well as the intrinsic dispersion of stream stars around the isochrone track.
For the former, we follow \citet{Shipp:2018} and fit the median $M_g$ error as a function of $M_g$ in PS1 to obtain the following function for observational uncertainties:
\begin{equation}
    M_{g,{\textrm{err}}} = 0.004 + \exp \left( \frac{M_g - 8.41}{1.10} \right)
\end{equation} 
For the latter, we choose an intrinsic width of 0.075 dex in $g-r$ based on visual inspection of the result.
This makes the total width $w_{g-r}$ of the selection region in $g-r$:
\begin{equation}
    w_{g-r}(M_g) = 0.075 + 2 \, M_{g,{\textrm{err}}}
\end{equation}
where we multiply $M_{g,{\textrm{err}}}$ by 2 to ensure we get all stars within 2$\sigma$ of the theoretical isochrone.
This is wide enough for us to be confident that we are almost complete for GD-1's main sequence stars, which we re-assess later.
Finally, we also select the horizontal branch region of the CMD space.
The specific cut we make is shown in the right panel of \figref{gaia_cuts}, along with the main sequence isochrone selection.
We do not include the red giant branch of the isochrone because the signal-to-noise (in number of GD-1 stars vs. number of background stars) in this region is low and it would decrease the purity of the sample significantly without much benefit.

\begin{figure}[t!]
    \centering
    \includegraphics[width=\linewidth]{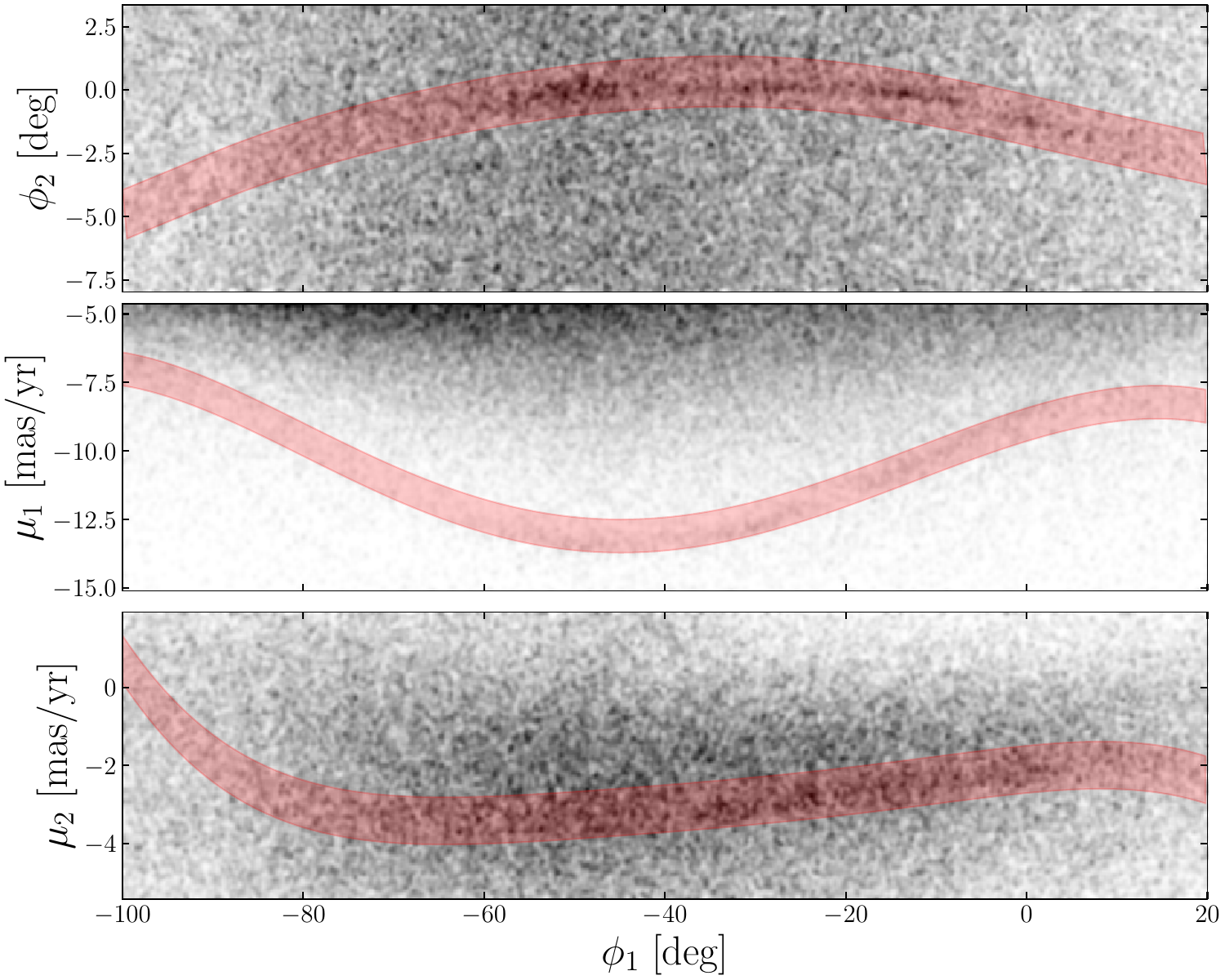}
\caption{Three different projections of the sample of stars used as input to the density model, all as a function of stream longitude $\phi_1$.
These stars pass the proper motion and CMD selections described in \secref{data} (and shown in \figref{gaia_cuts}).
The grayscale images show the number of stars in each bin, and the red shaded regions show the mean track and dispersion of the stream from \texttt{galstreams} \citep{Mateu:2023} that are used for initialization of the density model (described in \appref{initialization}).
\textbf{Top}: Sky positions of stars in stream longitude $\phi_1$ and latitude $\phi_2$ \citep{Koposov:10}.
\textbf{Middle}: Proper motion in $\phi_1$ for stars in the above sky region. The extent of the y-axis in this panel shows the selection in $\mu_1$ used to create the model input dataset.
\textbf{Bottom}: Proper motion in $\phi_2$ for stars in the above sky region. The extent of the y-axis in this panel shows the selection in $\mu_2$ used to create the model input dataset.}
\label{fig:gd1_data}
\end{figure}

We collect radial velocity measurements from several public spectroscopic surveys and data sets: \Gaia DR3 \citep{Gaia:2023}, APOGEE DR17 \citep{Majewski:2017, SDSSDR17}, SEGUE \citep{Yanny:2009}, DESI \citep{DESI:2023, DESI:2024}, and MMT/Hectochelle observations of stars targeted to be associated with GD-1 \citep{Bonaca:20b}.
We cross-match these datasets to our \Gaia--PS1 sample in the GD-1 sky footprint and use the radial velocities from these surveys as our $v_r$ measurements.
Out of our full \Gaia--PS1 sample, 16,610 stars have \Gaia RVs, 1,339 APOGEE, 16,175 SEGUE, 7,808 DESI, and 1,160 MMT/Hectochelle (with many overlapping between surveys).
When multiple measurements are available for a star, we use the inverse-variance weighted average of the measurements using their reported radial velocity uncertainties.
For stars with no radial velocity measurement, we assign $v_r = 0 \kms$ and $\sigma_{v_r} = 10^4 \kms$ (an arbitrarily large value meant to null out the contribution of stars with missing RVs to the likelihood).

\subsection{Model choices for GD-1} \label{sec:model_choices}

We summarize our model choices for the GD-1 background, stream, and off-track components in Table~\ref{table:gd1_choices} and below.
These choices aim to balance model flexibility with computational efficiency and interpretability.
The model is designed to be easily adaptable to other streams while capturing the key physical features we expect to see in GD-1 based on previous studies.

Our modeling framework uses cubic splines as a function of $\phi_1$ to represent smooth variations in the model parameters.
Cubic splines are piecewise functions that connect third-degree polynomial segments at specified points called ``knots'' to create a twice continuously differentiable curve.
In our implementation, the knot locations (in $\phi_1$) are fixed hyperparameters that determine the physical scales over which the model can detect structure: small separations between knots enable the model to capture small-scale features, while large separations smooth out the functional dependencies of parameters in the model.
We discuss our specific choices for knot placements for each parameter below.
For all components, we model the sky positions $(\phi_1, \phi_2)$, proper motions $(\mu_1, \mu_2)$, and radial velocity $v_r$.

\begin{table*}
\centering
{\renewcommand{\arraystretch}{1.2}
\begin{tabular*}{0.7\linewidth}{ @{\extracolsep{\fill}} c c c}
 Coordinate & Probability Distribution & Parameters\\
 \hline \hline
 \multicolumn{3}{c}{\textbf{Background}} \\
 $\phi_1$ & Gaussian mixture model (GMM) & $\{\alpha_k\}_{K_{\phi_1}}$, $\{\sigma_k\}_{K_{\phi_1}}$ \\
 $\phi_2$ & uniform & -- \\
 $\mu_1$ & mixture of two truncated normals & $\bs{\alpha}$, $m_1(\phi_1)$, $m_2(\phi_1)$, $\sigma_1(\phi_1)$, $\sigma_2(\phi_1)$ \\
 $\mu_2$ & mixture of two truncated normals & $\bs{\alpha}$, $m_1(\phi_1)$, $m_2(\phi_1)$, $\sigma_1(\phi_1)$, $\sigma_2(\phi_1)$ \\
 $v_r$ & mixture of two truncated normals & $\bs{\alpha}$, $m_1(\phi_1)$, $m_2(\phi_1)$, $\sigma_1(\phi_1)$, $\sigma_2(\phi_1)$\\
 \hline \hline
 \multicolumn{3}{c}{\textbf{Stream}} \\
 $\phi_1$  & GMM & $\{\alpha_k\}_{K_{\phi_1}}$, $\{\sigma_k\}_{K_{\phi_1}}$ \\
 $\phi_2$ & truncated normal & $m(\phi_1)$, $\sigma(\phi_1)$ \\
 $\mu_1$ & truncated normal & $m(\phi_1)$, $\sigma(\phi_1)$ \\
 $\mu_2$ & truncated normal & $m(\phi_1)$, $\sigma(\phi_1)$ \\
 $v_r$ & truncated normal & $m(\phi_1)$, $\sigma(\phi_1)$ \\
  \hline \hline
 \multicolumn{3}{c}{\textbf{Off-track}} \\
 $\phi_2, \phi_1$ & GMM & $\{\alpha_k\}_{K_{\phi_1}}$, $\{\sigma_k\}_{K_{\phi_1}}$ \\
 $\mu_1$ & truncated normal & $m(\phi_1)$, $\sigma(\phi_1)$ \\
 $\mu_2$ & truncated normal & $m(\phi_1)$, $\sigma(\phi_1)$ \\
 $v_r$ & truncated normal & $m(\phi_1)$, $\sigma(\phi_1)$ \\
\end{tabular*}}
\caption{Overview of our model choices for GD-1 background, stream, and off-track components.
For each of the Gaussian mixture model (GMM) components, we fix the means of the component distributions, but other parameters are allowed to vary (standard deviation $\sigma$ and mixture weights $\alpha$).
See \secref{model_choices} for further explanation.
}
\label{table:gd1_choices}
\end{table*}

\subsubsection{Stream Model Choices} \label{sec:stream_choices}

Our stream model represents the distribution of stream stars along the main track of GD-1 in density, position, and velocities.
We use a mixture of normals with fixed means separated by $5 \degree$ to model the density in stream longitude $\phi_1$, motivated by previous results \citep[\eg][]{Price-Whelan_Bonaca:2018} that show density fluctuations on $\sim10\degree$ scales.
We use truncated normal distributions to model the stream latitude, proper motion, and radial velocity distributions.
To control the parameters of these truncated normals, we use cubic splines with knots placed every $10\degree$ in $\phi_1$.
For the stream latitude and proper motion coordinates, these knots encompass the entire range of $-100\degree < \phi_1 < 20\degree$, but for the radial velocities, we reduce this range to the $\phi_1$ range with radial velocity measurements of probable stream member stars ($-82.29\degree < \phi_1 < 2.60\degree$).
We make these knot separation choices based on previous studies of GD-1 that show relatively smooth astrometric tracks.
A more detailed mathematical formulation of this model component is presented in Appendix~\ref{app:stream_math}.

To increase the speed of convergence when optimizing the full density model, it is important to initialize each component with reasonable starting parameter values.
We therefore first run a VI optimization of the stream component individually by fitting the model to a dataset only consisting of stars consistent with GD-1's sky position and proper motion tracks.
This selection is based on the PWB18 GD-1 tracks, and we show the selected regions in each space with red bands in \figref{gd1_data}.
We detail the initialization of this stream VI run in Appendix~\ref{app:initialization} and show the results in Figure~\ref{fig:stream_model}.
We note that this initialization is not perfect (clear from the $\phi_2-\phi_1$ residual in the bottom panel of Figure~\ref{fig:stream_model}), but it does provide a reasonable starting point for optimizing the full model.



\subsubsection{Background Model Choices} \label{sec:background_choices}

Our background model represents the smooth distribution of non-stream stars in the sky region around GD-1.
As with the stream model, we use a mixture of normals to model the density in stream longitude $\phi_1$.
Contrary to the stream model, however, we then assume a uniform distribution in stream latitude $\phi_2$, and use mixtures of truncated normal distributions to model the proper motion and radial velocity distributions.
We again use cubic splines to control the parameters of these truncated normal distributions, this time with knots placed every $40\degree$ in $\phi_1$.
This separation ensures that our background model cannot capture small-scale variations that might be associated with stream features.
The detailed mathematical formulation of this model component is presented in Appendix~\ref{app:bkg_math}.

As with the stream component, we run a VI optimization of the background component individually to improve the initialization of the full model.
We fit the background to the complement of the ``stream-only'' dataset we use to optimize the stream component (see \secref{stream_choices}).
We detail the initialization of this background VI run in \appref{initialization} and show the results in \figref{background_model}.
As with the stream model, this initialization is not perfect, but it provides a reasonable starting point for optimizing the full model.

\subsubsection{Off-track Model Choices} \label{sec:offtrack_choices}

Our off-track model represents the distribution of GD-1 stars that do not lie along the main stream track.
In other words, these are features in sky positions,
$(\phi_1, \phi_2)$, that deviate from the main stream track but have other kinematic observables suggesting a likely association with GD-1.
The key assumption for modeling this component is that the proper motions and radial velocities (referred to collectively in this subsection as ``kinematics'') of stream stars that migrate off the main stream track have kinematics similar to that of the main stream.
This assumption is supported by observed off-track structures (\eg GD-1's spur), the precision of our kinematic measurements, and the fact that strong interactions capable of significantly altering proper motions tend to scatter stars over a large area, rather than creating localized positional overdensities away from the main track.
Therefore, we adopt the same distributions to model the off-track kinematics as we do for the stream kinematics (\ie spline models controlling the parameters of truncated normal distributions along the stream).
However, while the stream distributions have relatively wide  priors (see Appendix~\ref{app:stream_math}), we define narrow priors for our off-track component to keep the kinematics similar to the main stream.
To achieve this, we first run a VI optimization of our model without the off-track component.
This ``background + stream'' model provides a (cubic spline) representation of the main stream's kinematic tracks, from which we define priors for the off-track component kinematics.

The primary element of the off-track component, however, is the density distribution in sky positions $(\phi_1, \phi_2)$.
We adopt a similar distribution as for the background and stream density models, but applied in the two dimensional (joint $p(\phi_1, \phi_2)$) space rather than as a conditional (\ie $p(\phi_2 \given \phi_1)$).
We use a 2D mixture of normals to model the off-track density on the sky, but with means fixed to regularly-spaced grid points in $\phi_1, \phi_2$.
This resulting grid of 2D normals allows us to recover localized overdensities associated with the stream anywhere in our data without assuming such overdensities exist.
We choose the separation of ``knots'' in this 2D space to be $3\degree$ in $\phi_1$ and $1\degree$ in $\phi_2$.
As with the other two components, we relegate the detailed mathematical formulation of this component to Appendix~\ref{app:offtrack_math}.

\subsection{VI choices}

We use the AutoNormal guide in numpyro to run VI on our model.
The AutoNormal guide approximates the posterior distribution by assuming each parameter is both independent and well-represented by a normal distribution.
This is not a correct assumption (\eg spline knot values should be correlated) but it should not bias our results significantly as it is a conservative approximation to the posterior \pdf.
We then optimize using the Adaptive Moment Estimation (Adam) optimizer with an adaptive learning rate and a maximum gradient for each step of 10.
The learning rate is made up of four segments of 2500 steps each.
All four segments consist of a learning rate that follows a cosine decay, starting at 0.1, 0.01, 0.001, and 0.0001 respectively, and each decreasing by one order of magnitude over the 2500 steps.
This ensures we search a large area of the parameter space at first, and then thoroughly explore the region near the best solution.
We initialize this VI run with a reasonably good guess for the stream and background and refer the reader back to Sections~\ref{sec:stream_choices}, \ref{sec:background_choices}, and \ref{sec:offtrack_choices} as well as Appendix~\ref{app:initialization} for details.
After the first VI run, we use the results to initialize a new run and repeat this process a few times until the ELBO stabilizes, signaling model convergence.

\section{Application to the GD-1 Stream: Results}
\label{sec:gd1_density_model}

\begin{figure*}[th!]
    \centering
    \includegraphics[width=\linewidth]{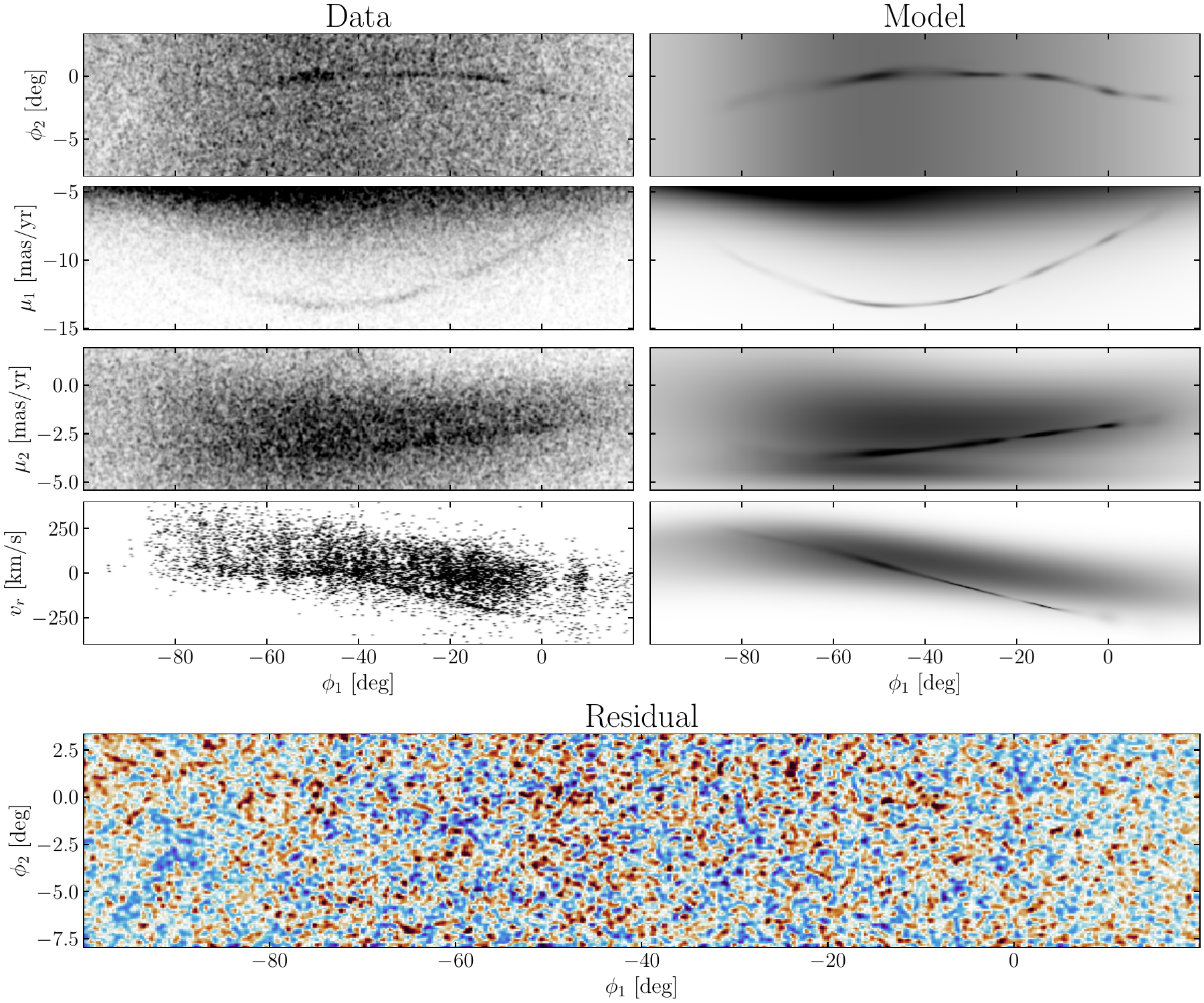}
    \caption{The complete model for GD-1. The left column shows the data, while the right column shows the model. The bottom panel shows the residual between the data and the model in position space. Each row displays a different space in which the model is created ($\phi_2 - \phi_1$, $\mu_1 - \phi_1$, $\mu_2 - \phi_1$, and $v_r - \phi_1$ from the first to the fourth row).}
\label{fig:full_model}
\end{figure*}

The result of our full model is shown in \figref{full_model}.
In the following sections we make use of two different types of results from VI.
First, since our VI run generates a posterior distribution for each parameter, we generate samples from these distributions to describe the uncertainty in each parameter's estimate.
While these distributions are important, we also want to be able to show the best-fitting parameters found by our model.
We do this by taking the mean value outputted for each parameter.
For simplicity, we henceforth refer to this parameter set as the best-fitting parameters.
In the subsections below, we detail inferred properties of the density model for GD-1 (e.g., off-track features) and derived quantities (e.g., total stellar mass).

\subsection{Inferred Properties} \label{sec:inferred_properties}

\subsubsection{Off-Track and Non-Gaussian Features} \label{sec:offtrack_results}

\begin{figure*}[th!]
    \centering
    \includegraphics[width=\linewidth]{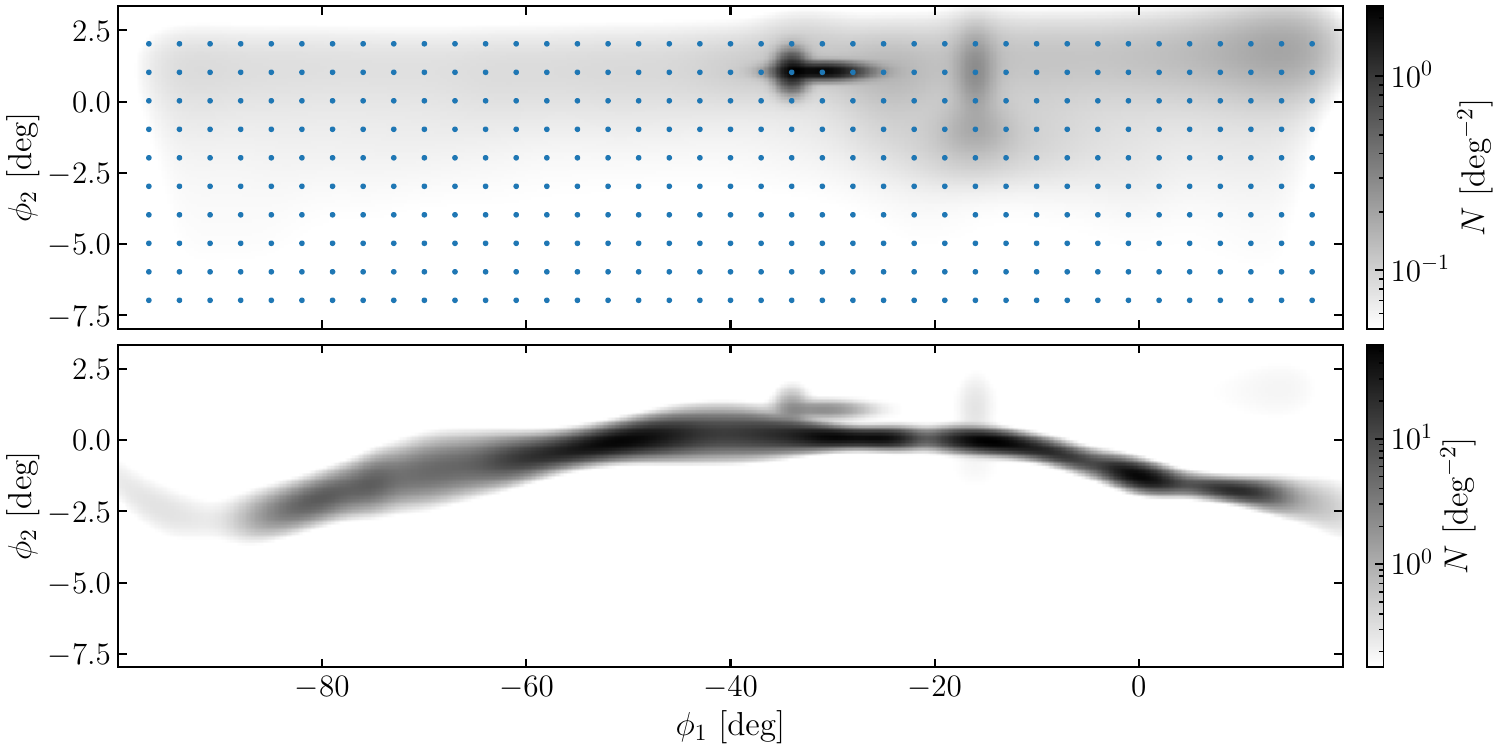}
\caption{\textit{Top}: The off-track model for GD-1 on a logarithmic color scale, as described in \secref{offtrack_choices}. The blue points represent the centers of our 2D normals used to create the off-track density.} We call special attention to the fact that we recover both the ``spur'' and the ``blob'' first seen in \citet{Bonaca:2019}.
\textit{Bottom}: The combined stream and off-track model in $\phi_1$--$\phi_2$ space with a logarithmic color scale so that the off-track components are visible.

\label{fig:offtrack_model}
\end{figure*}

\begin{figure*}[th!]
    \centering
    \includegraphics[width=0.95\linewidth]{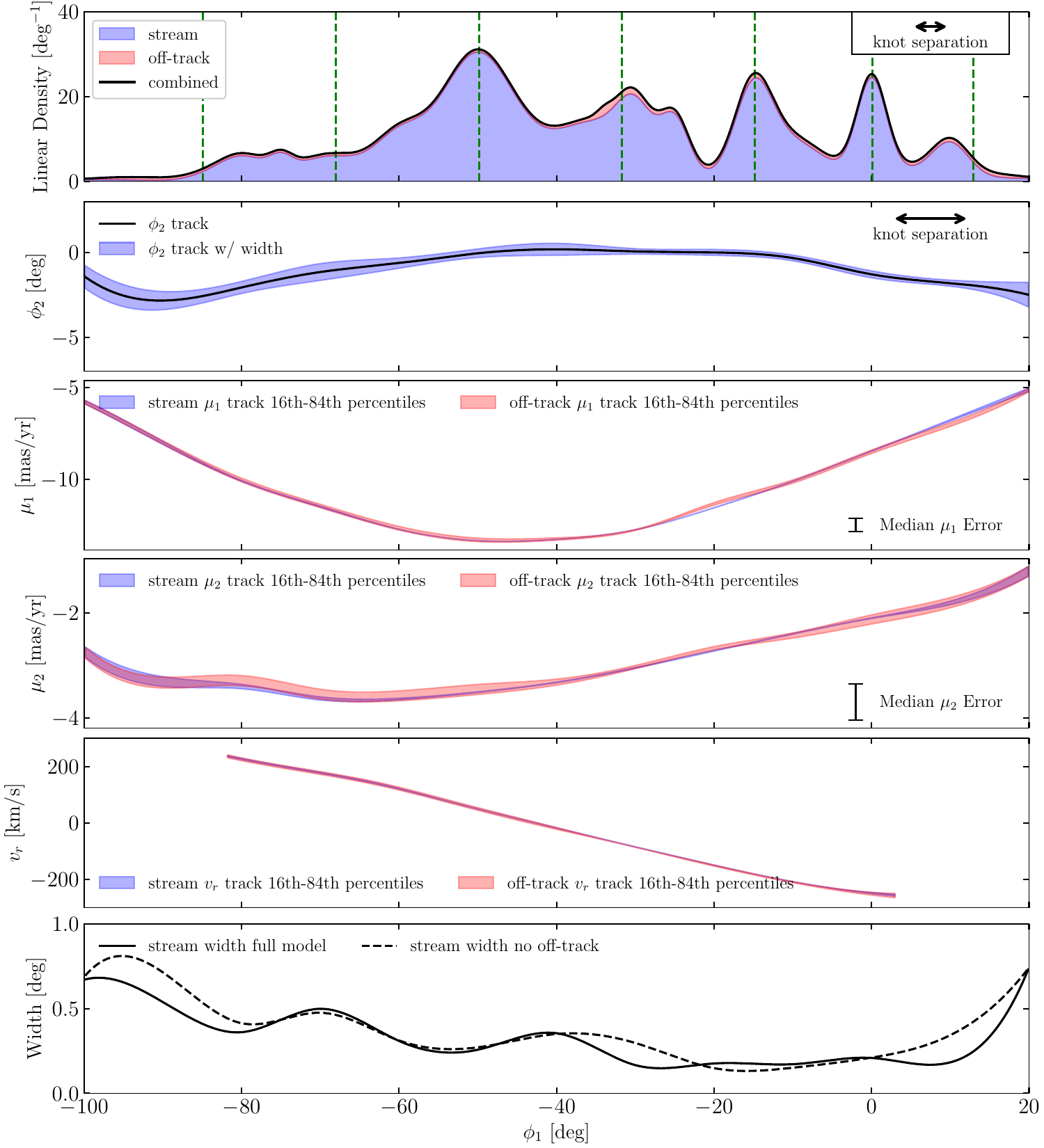}
\caption{Properties of the GD-1 stream as a function of $\phi_1$.
\textit{Top panel:} Linear density of GD-1. The blue shaded region represents the linear density on the main stream track and the red shaded region shows the linear density of the off-track components. The black line traces the combined linear density from both components. The green dashed lines show the locations of the density peaks expected from the best-fit epicycles (2.35 kpc apart). We indicate the knot separations for our stream and off-track $\phi_1$ Gaussian mixture model in the upper right corner of the panel.
\textit{Second panel:} Mean $\phi_2$ track (traced by the black line) and $1\sigma$ width (shown as the blue shaded region) of our GD-1 model's stream component. We indicate the knot separations for our stream and off-track velocity splines ($\mu_1, \mu_2$, and $v_r$) in the upper right corner of the panel.
\textit{Third panel:} Mean $\mu_1$ track (traced by the black line) and $1\sigma$ width (shown as the blue shaded region) of our GD-1 model's stream component. The red shaded region shows the $1\sigma$ width of the off-track $\mu_1$ track.
\textit{Fourth panel:} Same as the third panel but for $\mu_2$.
\textit{Fifth panel:} Same as the third panel but for $v_r$.
\textit{Bottom row:} The width of the main stream track (also represented by the red shaded region in the second panel). The solid curve traces the width of the main stream track when including the off-track component and the dashed line shows the width when only using background and stream components. This shows the importance of using an off-track component to characterize GD-1 properly.
In the third and fourth panel, we show the median proper motion errors of high probability GD-1 members ($\approx 0.35 \masyr$)
}
\label{fig:stream_properties}
\end{figure*}

Our main result is the recovery of off-track and non-Gaussian features that appear associated with the stream.
The most significant such feature is the ``spur'', first  identified by PWB18.
The spur appears prominently in both panels of \figref{offtrack_model} at $\phi_2 > 0 \degree$ and $-40 \degree < \phi_1 < -25 \degree$.
The model was not told that this feature existed and was not predisposed to find it in any way.
The fact that it is able to do so demonstrates its success in the major innovation our model contains over previous versions.

The other previously discovered overdensity (the ``blob''; $\phi_1 \approx -16$, $\phi_2<0$) is not as prominent in our results, although it is the off-track region with the highest density other than the spur.
Interestingly, the slight overdensity we do recover in this region is at both positive and negative $\phi_2$ values, in contrast to the blob in PWB18, which is exclusively below the stream.
This may be a reason for the relative faintness of this feature and we address this point further in \secref{offtrack_vel_offset}.


We then check whether our model identifies any additional off-track features.
We specifically seek overdensities with small-to-medium spatial scales that suggest a complicated origin.
In the top panel of \figref{stream_properties}, we show the linear density of the off-track component in the red shaded region.
This shows a low level of structure in the off-track component over the entire stream region.
Comparing this to the top panel of \figref{offtrack_model} suggests that all of this structure is extremely low surface brightness and not at all localized, although the off-track density becomes even less dense further away from the stream (also see \figref{high_prob_members}).
The most likely explanation for the general trend of higher off-track density at higher $\phi_2$ is that the background is not perfectly uniform in stream latitude, as we assumed.
A different, more intriguing explanation is that these stars constitute a ``cocoon'' around GD-1 \citep{Malhan:19, Valluri:25}.
By examining the high-probability off-track members, rather than surface density maps (see \secref{membership} and \figref{high_prob_members}) we do find most of these to lie within a couple degrees of the stream track for the region $-60<\phi_1<0$, similar to the cocoon claimed in \citet{Malhan:19}.
We do not find evidence that such a cocoon extends the entire length of the stream.

\subsubsection{Membership} \label{sec:membership}

\begin{figure*}[th!]
    \centering
    \includegraphics[width=\linewidth]{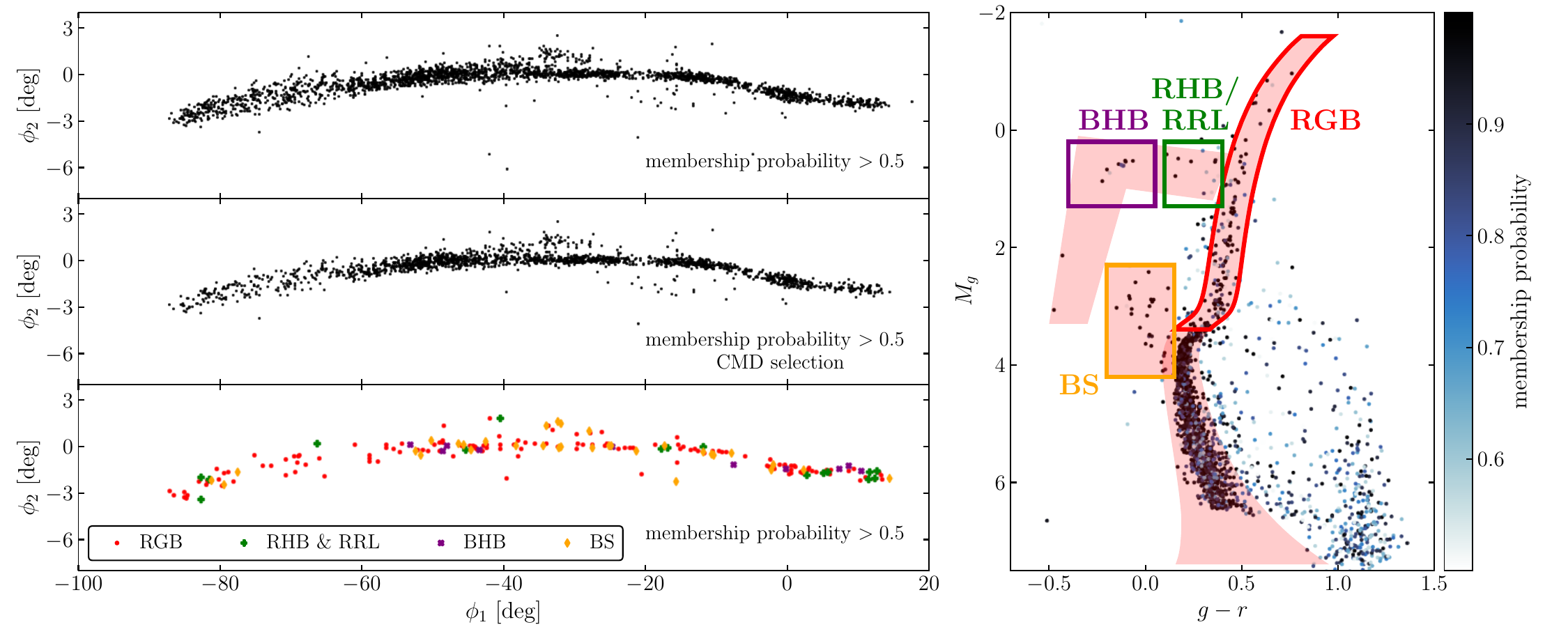}
\caption{High probability members of GD-1. \textit{Left:} Each panel shows an example cut one might make to select GD-1 members. The top panel shows all stars with a membership probability $>0.5$. The middle panel shows all stars with membership probability $>0.5$ that also pass a CMD cut (shown with the red shaded region on the right).
The bottom panel shows all non-main sequence stars with membership probability $>0.5$ that also pass a CMD cut, colored based on their location on the CMD diagram.
\textit{Right}: Distance-corrected color-magnitude diagram of stars with GD-1 membership probability $>0.5$. The color-scale shows the membership probability. The red shaded region represents the best-fit isochrone and horizontal branch of GD-1 based on a combination of prior studies and a fit to the CMD data (see \secref{data} for details). We also label with rectangles or polygons the approximate locations of blue horizontal branch (BHB) stars, red horizontal branch (RHB) and RR Lyrae (RRL) stars, blue stragglers (BS), and the red giant branch (RGB) in purple, green, orange, and red boxes, respectively.}
\label{fig:high_prob_members}
\end{figure*}

We calculate the probability that each star in the GD-1 region belongs to either the stream or off-track components of the model (vs. the background component).
We refer to this as the membership probability of the star, which we compute using the posterior predictive distribution of the model tracks.
In \figref{high_prob_members}, we show those stars with membership probabilities above 50\%.
In the right panel, we show the color--magnitude diagram of all stars with membership probability $>0.5$.
The red shaded region shows an example selection polygon around the best fit isochrone (same isochrone as used to make the original CMD selection in \secref{data}, but including the red giant branch, and the blue straggler population).
For most studies the best balance between purity and completeness will likely come from a combination of this CMD cut with the membership probabilities.
We show such a cut (membership probability $>0.5$ and passing CMD cut) in the middle left panel.
We henceforth refer to the 1689 stars which pass this cut as the high probability members of GD-1.
To complement this paper, we provide a catalog of stars in our original footprint with their \textit{Gaia} and PS1 astrometric and photometric properties as well as columns to help create pure or complete samples (membership probability and CMD mask).
We show three columns of this catalog for ten sample stars in Table~\ref{table:memb_tbl} and provide the rest of the catalog as a machine readable file on Zenodo, available at 10.5281/zenodo.15428120.

\begin{table}[h]
\centering
\begin{tabular}{c c c}
\textit{Gaia} source ID & memb\_prob & pass\_CMD\_cut \\
\hline
582446459246191232 & 0.9584 & True \\
582459584666428032 & 0.9371 & True \\
582461371372666752 & 0.8824 & False \\
576748068435698048 & 2.254e-93 & True \\
576748244530543488 & 1.155e-70 & False \\
576750301818727936 & 1.040e-05 & False \\
576760064280542592 & 5.345e-63 & False \\
576760300502579328 & 4.498e-60 & True \\
576760334862325504 & 6.697e-07 & False \\
576760437941539712 & 1.607e-17 & False \\
\end{tabular}
\caption{10 abridged sample rows from our membership probability table. For readability, the membership probabilities (second column) are truncated to four significant figures. As an example, the first three are stars with high membership probabilities and the others have low membership probabilities. The full table has many additional columns, with all relevant \textit{Gaia} astrometric and PS1 photometric data.}
\label{table:memb_tbl}
\end{table}

\subsubsection{GD-1 Linear Density} \label{sec:density_track}

As in previous studies, GD-1 appears to extend from $\phi_1\sim-90$ to $\phi_1\sim10$.
Outside of this region, the main stream track density approaches 0.
In between, we recover density fluctuations along the stream, including four prominent gaps and four peaks.
The gaps occur at $\phi_1 \approx -40, -20, -5, 5 \degree$ while the density peaks lie at $\phi_1 \approx -50,-33,-15,0, 10$ (seen best in the top panel of \figref{stream_properties}).
Fitting a power spectrum to the density profile, we recover the typical scale of density fluctuations to be 2.35 \kpc.
This is within $2\sigma$ of the $2.64 \pm 0.18 \kpc$ fluctuation scale reported in \citet{Ibata:20}, which they attribute to epicyclic motion.\footnote{In Figure 22 of \citet{Ibata:24}, the authors show the epicyclic separation derived in \citet{Ibata:20}. Although they have improved data in the more recent paper, they do not re-perform this analysis and a visual inspection suggests that their initial result may have been a slight overestimate, and an updated result would be more in line with our epicyclic separation of 2.35 \kpc.}
This theory is compelling given how evenly spaced many of the density peaks appear to be (green dashed vertical lines in the top panel of \figref{stream_properties}.
The amplitude of these fluctuations in the central stream region ($-60 \degree < \phi_1 < 0 \degree$) is significant, with density peaks having $\approx2-8$ times the linear density as the gaps.

\subsubsection{GD-1 Track and Width} \label{sec:track and width}

Apart from the linear density, our model also infers $\phi_2$ and proper motion tracks along the stream (middle three panels), together with the widths of those tracks (bottom panel).
GD-1's track in $\phi_2$, as shown in the second panel of \figref{stream_properties}, does not have any major features of interest; it is smooth and contains no small scale variations.
The uniformity of the stream track is naively expected from simple models of stellar streams.
However, some streams, such as ATLAS-Aliqa Uma and Phoenix, have $\phi_2$--track discontinuities or small scale wiggles \citep{Li:2021, Tavangar:22}, most likely caused by a past interaction or some complicated history.
Since the spur in GD-1 indicates it may have had a significant interaction within the last \Gyr, it is notable that the cause of that feature did not also impact the stream track on small scales.
We note, however, that the lack of sharp kinks or small scale wiggles does not necessarily imply that the spur-causing interaction had no effect on the track.
It is plausible that such an interaction could uniformly shift large segments of the stream off its original orbital path.
This could happen if the perturber impacting the stream is on a close-to-parallel orbit with the stream stars, such that it affects an extended region of the stream.

The width of GD-1 is shown in the bottom panel of \figref{stream_properties}.
Similarly to the track, we see no small scale changes in the width.
However, the width track still contains two interesting features: the gradual increase in stream width at $\phi_1 \lesssim -30$ and its oscillation in this region.
One advantage of having an off-track model is that we can test whether these features are real or caused by an extra component such as the spur.
To do this, we compare the width of our full GD-1 model with the stream width in a model with no off-track component.
The two are nearly identical except in the outskirts and in the spur region ($-40 \lesssim \phi_1 \lesssim -20$), where the stream$+$background model is slightly wider in an attempt to fit the spur without an off-track component.
This mischaracterization of the spur region reinforces the need for an off-track component to ensure unbiased studies of stream properties.
The identical result at lower $\phi_1$ also confirms that the gradual width increase and oscillation for $\phi_1 \lesssim -30$ is real.
Since the stream is more distant in this region (lower $\phi_1$), this increase in angular width also corresponds to an increase in physical (\ie \kpc) width.

\subsubsection{GD-1 Velocity Tracks}
The kinematic tracks and widths (shown in the third, fourth, and fifth panels of \figref{stream_properties}) are relatively simple, in the sense that these properties vary smoothly along the stream with no small angular scale features.
They also have good agreement with the previous literature as documented in \texttt{galstreams} \citep{Ibata:21, Price-Whelan_Bonaca:2018}.

\begin{figure}[t!]
    \centering
    \includegraphics[width=\linewidth]{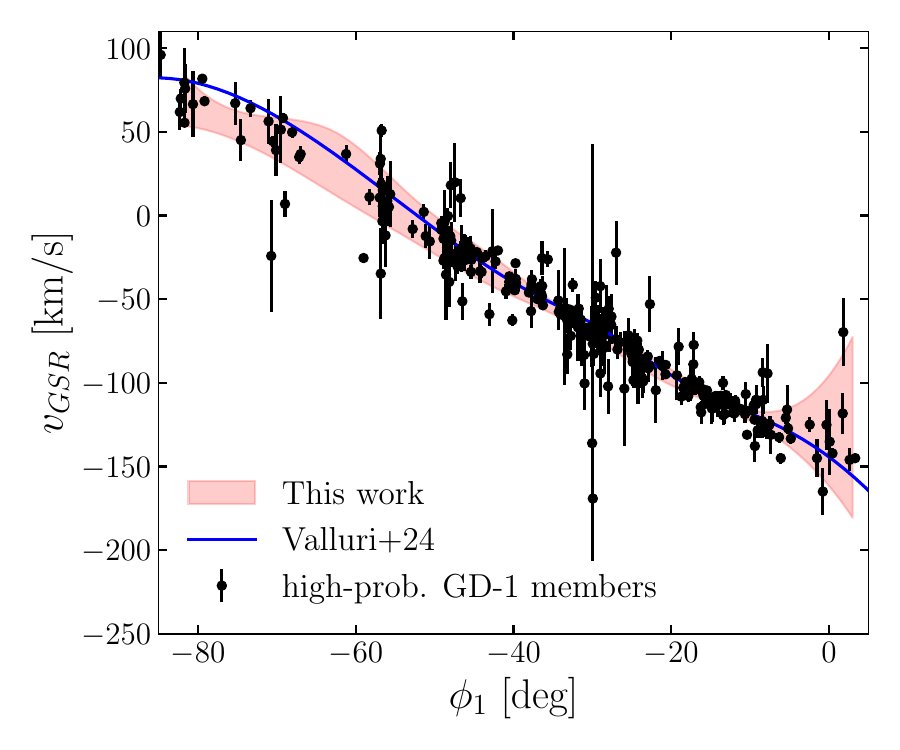}
\caption{High probability (membership probability $> 0.5$ and passing the CMD cut) members of GD-1 with radial velocity measurements, converted to the Galactic Standard of Rest frame (GSR). We show the radial velocity track from our model (including its width) in red and the best-fit radial velocity track from \citet{Valluri:25} in blue.}
\label{fig:high_prob_rv}
\end{figure}

One of the main results of this work is an up-to-date catalog of radial velocities for GD-1 members.
We find 353 high probability members (membership probability $> 0.5$ and passing the CMD cut) with radial velocity measurements in our sample, which we show in \figref{high_prob_rv}.
We compare these members to the members with radial velocity measurements recovered by two recent analyses: \citet{Ibata:24} and \citet{Valluri:25}.
\citet{Ibata:24} find 323 GD-1 members with radial velocities.
We have 6D phase space measurements for 249 of these 323 in our work and find that 221 of those are likely to be GD-1 members ($\sim$89\% overlap among commonly analyzed stars).
\citet{Valluri:25} find 145 likely members, all of which are also in our sample with full 6D phase space information.
We find that 120 of them pass our membership cut ($\sim$83\% overlap).
Our radial velocity track, including its width, is over-plotted in the red shaded region.
We compare this to the best fit radial velocity curve from \citet{Valluri:25} (shown in blue). 

\subsubsection{Metallicity and \emph{[$\alpha$/Fe]}} \label{abundances}

\begin{figure*}[th!]
    \centering
    \includegraphics[width=\linewidth]{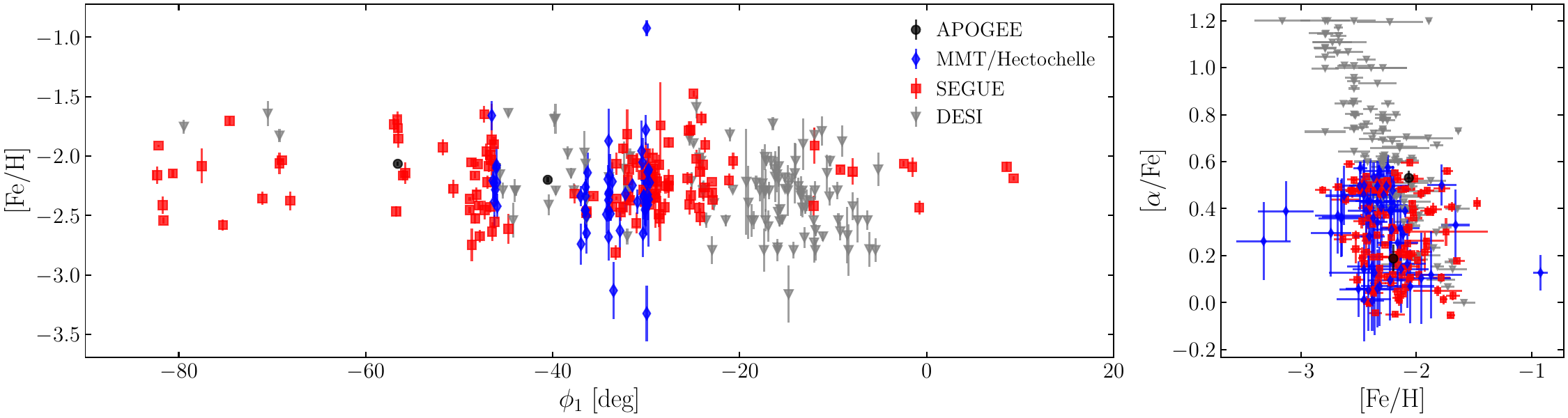}
\caption{\textit{Left:} [Fe/H] \vs $\phi_1$ of likely GD-1 members (membership probability $>0.5$ and passing the CMD cut) for which we have spectroscopic measurements. \textit{Right:} [$\alpha$/Fe] \vs [Fe/H] of the same likely GD-1 members. These abundances are taken from APOGEE DR17 (black circles), MMT/Hectochelle (blue diamonds), SEGUE (red crosses), and DESI (gray triangles).}
\label{fig:chemistry}
\end{figure*}

The spectroscopic datasets used to get radial velocities of stars in the GD-1 region also provide abundances\footnote{For any stars that are in more than one of these datasets, we use the inverse-variance weighted average abundance values.}.
For the DESI dataset, we follow the guidance of \citet{DESI:2023} and add 0.21 dex to their [Fe/H] measurements to put them in line with APOGEE.
We can then examine the metallicities and $\alpha$ abundances of high-probability members (membership probability $>0.5$ and passing the CMD cut).
We also remove stars with measured metallicities that differ by more than $1$ dex from the literature value of $\approx -2.5$ dex \citep{Ibata:24, Valluri:25}.
This yields an inverse-variance weighted mean [Fe/H] of -2.19.
We then examine whether there is a metallicity gradient along the stream by plotting [Fe/H] vs $\phi_1$ as shown in the left panel of \figref{chemistry}.
Combining all the datasets, we recover a metallicity gradient of $-0.0033$ dex/deg. 
One caveat to the combined gradient is that there may be systematic differences between the survey abundance measurements that can bias these results.
However, if this gradient is real, it is interesting because it is not clear how this would happen.

We also look at the $\alpha$-abundance measurements for GD-1 stars, plotting all high probability members on the [$\alpha$/Fe] vs [Fe/H] plane (shown in the right panel of \figref{chemistry}.
We recover a weighted average [$\alpha$/Fe] abundance of 0.67 dex when combining all datasets.
This result remains if we only use DESI data but with only SEGUE data, we recover a lower [$\alpha$/Fe] of 0.30 dex. 
We recommend using this lower value as the true GD-1 [$\alpha$/Fe] abundance since the DESI data suggests an unphysically large range in addition to their extremely low uncertainties.

\subsubsection{Velocity Dispersion} \label{sec:dispersion}

Absent of any external heating, velocity dispersion along a stream is expected to (locally) decrease as the stream spreads out spatially \citep{Helmi:1999}.
However, in the presence of massive perturbers (e.g., dark matter subhalos, dwarf galaxies, etc.), the velocity dispersion of a stellar stream is indicative of its past heating from external structures within the halo.
This is because we can predict the intrinsic velocity dispersion, which depends on the progenitor and orbital properties.
A more massive progenitor will have a higher intrinsic velocity dispersion, as will a stream closer to apocenter.
The actual stream velocity dispersion then also depends on heating the stream experiences from encounters with external objects.
Each encounter will increase the velocity dispersion.
By comparing the expected intrinsic velocity dispersion with the observed one, we can estimate the population of subhalos in the Milky Way halo.

\begin{figure}[t]
    \centering
\includegraphics[width=\linewidth]{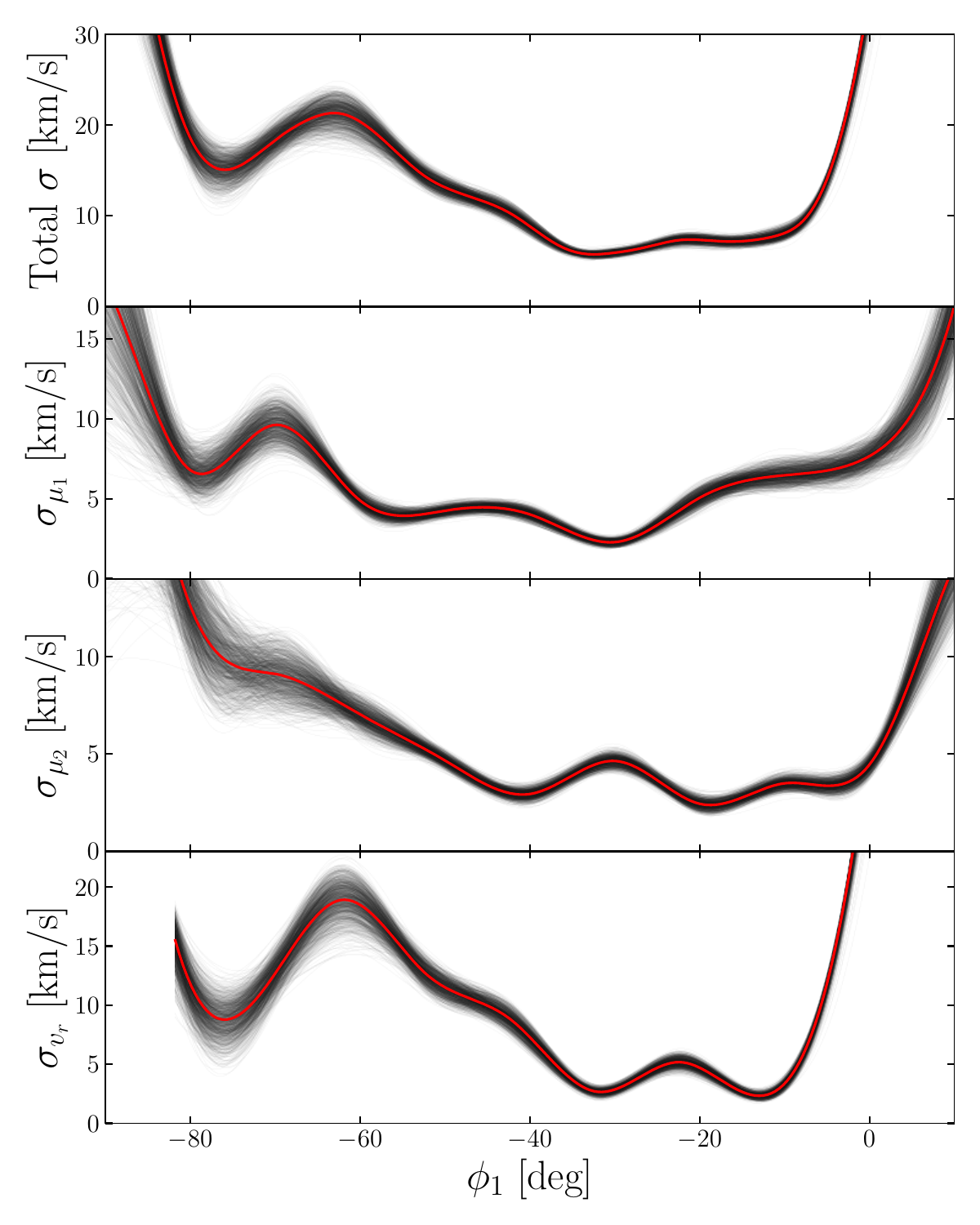}
    \caption{The black lines show 1000 samples of the total velocity dispersion of GD-1 as well as the velocity dispersion in each coordinate, as a function of $\phi_1$.
    The red line shows our best-fit model results.}
\label{fig:vel_dispersion}
\end{figure}

We show the velocity dispersion derived from our model in \figref{vel_dispersion}, where each black line represents one sample from the posterior distribution and the red line is the result from our best-fit model.
The proper motion dispersions (converted into velocity using the adopted distance track from \citealt{Valluri:25}) are relatively constant at $\approx 3$--$5 \kms$ for the central region of the stream ($-65 \degree <\phi_1 < 5 \degree$).
It is surprising that, given the average proper motion uncertainty for GD-1 stars ($\approx 0.35 \masyr$ or $\approx 12.5 \kms$ in each dimension), the model converges to a finite value less than this uncertainty, rather than be consistent with zero.
This is tentatively exciting but it is also possible that mistakes in calibration or error estimates created this effect (any systematic uncertainties will be interpreted as intrinsic velocity dispersion).
We neglect the uncertainty on the distance track, which will impact the uncertainties on these velocity dispersion profiles, but this should only add an additional $\sim 10\%$ uncertainty to the inferred dispersions.

The radial velocity dispersion is more varied, with $\sigma_{v_r} \lesssim 5 \kms$ for $-40 \lesssim \phi_1 \lesssim -10$ but higher dispersions elsewhere.
The minimum occurs at two locations: $\phi_1 \approx -15\degree, -32\degree$ .
Under CDM, it is reasonable to expect the radial velocity dispersion to be smallest at the progenitor location and larger at the stream edges, so this result could give another clue towards the progenitor location.
However, we caution that because we combined multiple surveys to create our radial velocity dataset: offsets between these surveys or inaccurately reported uncertainties could bias our inferred $v_r$ dispersion profile.
If instead of the model results we use the 353 high-probability ($>50\%$) members with radial velocity measurements and fit a 3rd degree polynomial to them as a function of $\phi_1$, we recover a velocity dispersion of 2.35, slightly lower than \citet{Valluri:25} but in good agreement with \citet{Gialluca:21} and \citet{Nibauer:24}.
This value is consistent with expectations from CDM and a dark matter model with more compact subhalos ($0.5 \times r_s^{\textrm{CDM}}(M)$) but disfavors lower numbers of subhalos like one would expect in WDM or fuzzy DM \citep{Nibauer:24}.

\subsection{Derived GD-1 Properties}

\subsubsection{Off-track Velocity Offset} \label{sec:offtrack_vel_offset}

\begin{figure*}[t!]
    \centering
    \includegraphics[width=\textwidth]{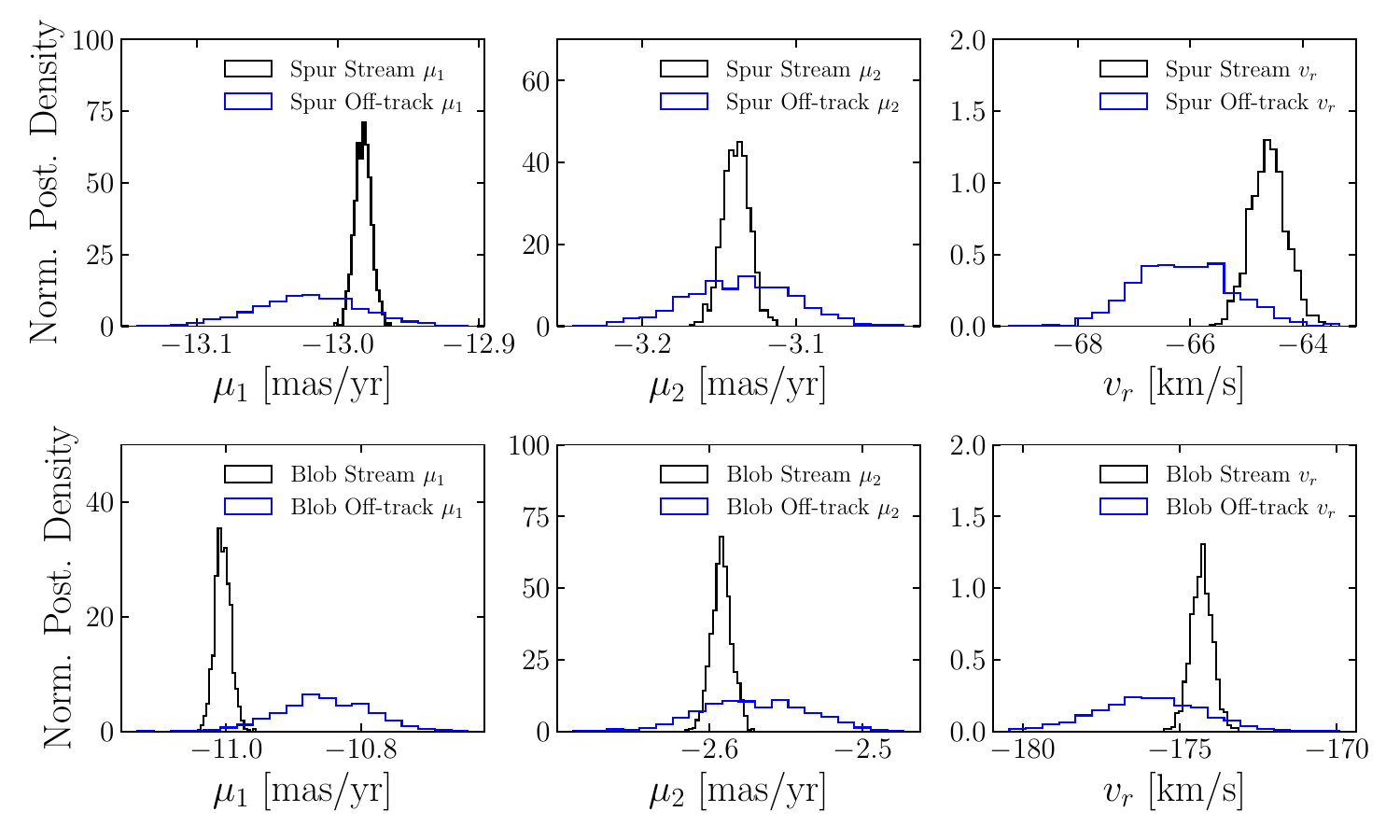}
\caption{Histogram of the mean proper motion and radial velocities for the stream and off-track components from our model posterior near the spur (top row) and the blob (bottom row).
We obtain this by creating a proper motion and radial velocity spline from each posterior sample and evaluating $\mu_1$ and $\mu_2$ at $\phi_1\approx -33$ (spur) and $\phi_1 \approx -16$ (blob). 
The off-track distributions are consistent with the main tracks' except for the radial velocity of the spur and the $\mu_1$ of the blob.}
\label{fig:offtrack_offsets}
\end{figure*}

In theory, in order for a star to drift away from the main stream track, it must have a different velocity from the main stream.
If it does not, it has no way of distancing itself from other stream stars.
Therefore, we check the velocity difference between the spur and the main stream track at the equivalent $\phi_1$.
We do so by evaluating the stream and off-track $\mu_1$, $\mu_2$, and $v_r$ cubic splines at the $\phi_1$ location of the spur ($\phi_1=-33\degree$). 
We find the mean proper motions of the spur to be entirely consistent with the main stream's velocity.
However, the spur does appear to have a slightly lower radial velocity than its main track counterpart.
We show these off-track and main stream velocity distributions in the top row of \figref{offtrack_offsets}, where we plot histograms of the inferred velocities at the spur location from our posterior samples. 

We then check the other previously discovered off-track feature (the blob).
and show the distributions of velocities in the blob region in the bottom row of \figref{offtrack_offsets}.
In contrast to the spur, the main difference for the blob is in $\mu_1$, where the off-track component has a higher $\mu_1$ than the main stream.
This difference could explain why we recover a less robust blob than PWB18.
By itself, a velocity difference should have no effect on the model's ability to recover this overdensity, since we allow off-track proper motions to differ slightly from the main stream.
However, our current model still recovers off-track proper motions as a function of $\phi_1$ only, meaning the proper motion distributions are assumed to be independent of $\phi_2$.
Therefore, if the off-track features above and below the main stream in the blob region have separate proper motion distributions (\eg there is a $\mu_1$ gradient along $\phi_2$ at $\phi_1\approx -16$) the model will have difficulty recovering the off-track $\mu_1$ distribution accurately.
This gradient theory is supported by the fact that if we run a version of our model with no proper motion errors (\ie no deconvolution and therefore broader proper motion distributions) we obtain a prominent blob for $\phi_2 < 0$ very similar to PWB18.
In future work, we will address this limitation.

\subsubsection{Stellar Mass} \label{sec:mass}
From our list of probable members, we can estimate both the initial and current stellar mass of GD-1.
We assume a Kroupa \citep{Kroupa:01, Kroupa:02} initial mass function (IMF) for the GD-1 progenitor and leverage the fact that we are complete for \Gaia $G<20.5$.
Specifically, the Kroupa IMF allows us to calculate that 6.1\% of the initial mass currently remains on part of the observable and complete  part of the main sequence ($3.2 \magn \lesssim M_g \lesssim 5.9 \magn$).
We detect 1278 stars with high membership probability ($>50\%$) on this part of the main sequence, and the sum of their masses, assuming a Kroupa IMF is $\approx 941 \Msun$.
This means that, under these assumptions, the initial mass of the GD-1 progenitor was $1.53 \times 10^4 \Msun$.
This is a factor of $\approx 3$ lower than the $5\times10^4$ value often assumed when creating dynamical models of GD-1 \citep[\eg][]{Ibata:24}.
We note that if our model does not recover the full extent of the stream, our recovered initial mass will necessarily be lower than the true initial mass. 
At both edges of the stream, GD-1's proper motions are less distinct from the background (see Figure~\ref{fig:full_model}) and the stream is therefore more difficult to recover.
Our model is designed to be flexible enough to account for this increased level of contamination but we leave open the possibility that it does not recover the full extent of the stream, in which case our mass estimate would be too low.

For the current mass, we can use a similar idea.
By the same logic as above, $\approx 36$\% of the initial mass of GD-1 remains undetected.
Combining this with the observed main sequence, we deduce that the total current main sequence mass of GD-1 is $\approx 6484 \Msun$.
To account for the mass of the stars brighter than the main sequence, we rely on the stellar masses from the best Dotter \citep{Dotter:2016} isochrone fit to the GD-1 data (discussed in \secref{data}).
Since all these stars have recently evolved off the main sequence, they all have similar masses $M \approx 0.82 \Msun$.
Since we recover 191 stars with high membership probability which pass the CMD cut in this region and we know we are complete there, we now have the mass of every part of the current GD-1 stellar population.
Summing these masses yields a current GD-1 mass of $\approx 6645 \Msun$, approximately 43\% of its initial mass.

\subsubsection{GD-1 Progenitor} \label{sec:progenitor}

The GD-1 stream has no known surviving progenitor and is therefore assumed to have been fully disrupted.
If the progenitor recently disrupted, GD-1's non-uniform mass loss means it may leave behind an observable density gap \citep{Webb:19}.
Such a signature exists at several locations in GD-1.
Based on these features, other studies have suggested final progenitor locations of $\phi_1 \approx -40, -30, -20$ \citep{Carlberg:13, Webb:19, Ibata:24}.

The leading and trailing arms of streams are stripped from different sides of their progenitor clusters.
Therefore, another possible signature of a progenitor is a sharp S-shape in the stream track on the sky, as seen in Palomar 5 \citep[\eg][]{Bonaca:20a}, provided that this feature does not occur along the line of sight.
We do not find such a feature in our inferred stream model, with the caveat that the large separation of knots in the $\phi_2$ stream track would not allow us to recover a localized (degree-scale) change in the track.
However, the off-track component would be able to detect such a feature and it does not appear to have done so.
The only part of the stream with a gap and an apparent change in the concavity of the stream track is at $\phi_1 \approx -6\degree$.
We therefore do not find any additional kinematic evidence of the final progenitor location.

Another uncertainty with GD-1 is the nature of the progenitor (\ie whether it was a star cluster or low-mass dwarf galaxy).
Based on its stellar mass, low velocity dispersion, low metallicity, and small (unmeasured) spread in iron abundance, it is likely that GD-1 was a star cluster \citep[e.g.,][]{Bonaca:20b}.
Here, we find a number of horizontal branch (HB) and blue straggler (BS) stars associated with the stream.
The ratio of the number of HB stars to BS stars has been shown to correlate with the type of progenitor system \citep{Momany:2014, Deason:2015}.
For GD-1, we find $N_{\textrm{BS}} / N_{\textrm{HB}} = 32/26 \approx 1.23$ with a total absolute magnitude $M_V \approx -4.7$.
This unfortunately places GD-1 in a region consistent with both a globular cluster and low-mass dwarf galaxy progenitor, meaning this test is unable to resolve conclusively the nature of GD-1's progenitor.

\section{Discussion} \label{sec:discussion}

We now place our work in the context of the current state of the field and look ahead to future improvements.
We have largely built on previous work that uses cubic spline modeling methods, but add the key addition of an off-track component, giving us the flexibility to characterize known inhomogeneities and discover new ones.
This has worked well when applied to GD-1 but we anticipate the possibility to needing to tweak certain aspects of the model when it is applied to other streams which will be more difficult to characterize.
Our model currently has a number of limitations that we outline in \secref{future}, the most notable of which is that we currently make a CMD cut as opposed to modeling that space in the same way we do the astrometry.
We aim to make this improvement, among others, in future iterations of this model.

\subsection{Comparison to Past Density Models} \label{sec:comparison}

Our modeling framework builds upon past stellar stream density modeling methods.
Previous analyses of stream density variations used \texttt{STAN} \citep{Carpenter:17} for statistical modeling and efficient probabilistic sampling \citep{Erkal:2017, Koposov:2019, Li:2021, Tavangar:22, Ferguson:2022} or custom statistical modeling methods \citep[e.g.,][]{Bonaca:20a}.
These past efforts have revealed important stream features, like the track discontinuity in ATLAS-Aliqa Uma \citep{Li:2021}, varying density fluctuation scales between streams \citep{Tavangar:22}, and variations in the track and width of individual streams \citep[e.g.,][]{Bonaca:20a, Ferguson:2022}.
\citet{Patrick:2022} then analyzed 13 Milky Way streams to provide a population-level view of stream density structures.
Given that we now know of almost 150 streams in the Milky Way \citep{Bonaca:25}, it is clear that robust, automated, and efficient stream density modeling methods will be a critical tool for understanding this population.

Our framework builds upon these past methods by providing a more generalizable and accessible tool for stream density modeling.
In particular, our implementation is in Python and uses \texttt{jax} \citep{jax:18} and \texttt{numpyro} \citep{Bingham:19, Phan:19} for gradient-based optimization and flexible specification of probabilistic models.
We also provide a mechanism for modeling off-track density features in streams, which will be important for understanding the full extent of a stream and a full accounting of its member stars.
Our approach maintains computational efficiency while adding critical functionality for modeling complex stream morphology, as demonstrated in our application to GD-1.

\subsection{Future Improvements and Extensions} \label{sec:future}

Our model's performance on GD-1 is very promising for its application to other streams.
In the near future, this model will fold into ongoing work by the Community Atlas of Tidal Streams (CATS) collaboration.
CATS' objective is to build on the \texttt{galstreams} library by creating the tools necessary to perform analyses using multiple streams.
This involves developing analysis tools that are transferable from stream to stream, allowing easy comparisons.
CATS is also working on making easy-to-use selection boxes in different spaces for various streams, allowing users to make different selections of probable members depending on their use cases.
Our analysis will provide an improvement on such binary selections by outputting membership probabilities and a density model.

Despite this model's success, we see multiple areas of future improvement.
The most consequential change is to incorporate the CMD as part of the mixture model.
We remind the reader that in our work, photometric information is only used to create the dataset that our model uses as input.
However, since stream stars should lie along an isochrone, the color--magnitude distribution of stars in this field should be extremely informative for discovering and characterizing streams.
An improved model should make use of this additional information and add CMD modeling to the framework we have built here.
This will be crucial as we discover new streams in deep photometric surveys such as the Legacy Survey of Space and Time (LSST).
These streams will have limited proper motion information, especially below the main sequence turnoff, where we expect to find the majority of its members.
Therefore, characterization of those streams will rely heavily on CMD modeling.

CMD modeling is challenging because, in contrast to the astrometric background, the photometric background is too complex to model with simple analytic distributions.
This makes it difficult to model the isochrone overdensity lying atop the background.
There are a couple ways to tackle this challenge.
\citet{Patrick:2022} split the CMD into pixels (similar to our off-track cells) and modeled the background CMD as a discrete probability distribution of a star belonging to a certain pixel.
They were interested in modeling some streams without deep astrometric data, so it was critical for them to develop a successful CMD modeling method.
More recently, \citet{Starkman:2024} used normalizing flows to learn the background distribution of the CMD.
Normalizing flows provide flexible probability density descriptions by transforming samples from simple distribution (\eg Gaussian), to a more complex one \citep{Tabak:10, Rippel:13, Jimenez-Rezende:15}.
They combine this with an astrometric model and demonstrate that their method works well on the GD-1 and Palomar 5 streams.
Given the success of both these methods, we plan to allow future versions of our code to allow for easy inclusion of various CMD modeling techniques.

Once we model the color--magnitude distribution along with the kinematics, another important improvement will be to incorporate the survey completeness in our density modeling.
In any observational survey, the probability of detecting a star depends on a variety of factors, including its position and magnitude.
These selection effects can affect the density of certain types of stars on the sky, creating the illusion of stream features which do not truly exist.
Here, we have simply made a \Gaia $G$ cut to ensure there is no variation of the selection function with sky position.
This limits us because we are not using fainter stars on the main sequence which could help us characterize the stream.
This is an active area of research as shown by recent papers modeling the \Gaia selection function \citep[\eg][]{Cantat-Gaudin:23}.

There are also a couple smaller improvements we would like to implement.
For instance, when using a spline model, one must specify the number and placement of knots used to create the model.
The spacing should be related to the scale on which one expects to see structure in the object of interest.
For streams, this scale can vary by about an order of magnitude ($\approx 1 \degree$--$10 \degree$) \citep{Tavangar:22}.
Often, researchers choose the knot spacing by eye, simply trying to find the middle ground between under- and over-fitting.
In our case, as discussed in \secref{model_choices}, we used prior studies of GD-1 \citep[e.g.][]{Bonaca:2019} to approximate a knot spacing and ensured that small differences to this spacing did not significantly alter the model output.
Finding that it did not, we chose not to expend the significant computational expense required to create a full optimization algorithm for the knot spacing.
However, a more robust version of this model would include such an algorithm.
This will be useful for all streams, particularly those for which we do not yet know the density variation physical scales.


\section{Summary and Conclusions} \label{sec:conclusion}
We summarize our paper as follows:
\begin{itemize}
    \smallskip
    \item We present a new method for modeling the kinematic density distributions of stellar streams. Our method recovers the astrometric tracks of streams, the density of streams along their length, and the density of any off-track or non-Gaussian features associated with the streams. This framework lays the basis for uniformly inferring robust models of stellar streams to use as inputs to obtain tight constraints on the Galactic mass distribution and the nature of dark matter.
    \smallskip
    \item We apply our model to the GD-1 stream, which we know contains off-track features, to test its functionality. We recover similar results to previous efforts in terms of the astrometric tracks. We also recover previously-identified features (the spur and the blob) but provide more robust membership probabilities for stars associated with these substructures. In addition, the off-track component of our model indicates that there may be other diffuse low surface brightness GD-1 structure particularly in the region $-40\degree < \phi_1 < 0\degree$.
    \smallskip
    \item With our model for GD-1, it is straightforward to calculate membership probabilities for all stars in the region. We present a catalog of 466571 stars, 1689 of which have membership probability $> 0.5$ and pass our CMD cut.
    \smallskip
    \item Our GD-1 model results suggest an initial stellar mass of $1.53 \times 10^4 \Msun$, an average velocity dispersion along the main stream track of $\approx 6$--$15 \kms$.
    \smallskip
    \item We find the spur to have a slightly higher radial velocity than the main stream at the same $\phi_1$ while the other two velocity components are statistically consistent. Additionally, the blob is offset from its main stream velocity in $\mu_1$, while the other two components are statistically consistent.
    \smallskip
\end{itemize}

This contribution joins a number of recent analyses that define new methods for modeling the density of streams.
While it is useful to improve the methods we use for stream modeling, major scientific results will only come from a combined analysis of dozens of streams.
Therefore, we emphasize that once we have incorporated the improvements mentioned in \secref{future}, a broader application of our modeling framework is critical.
These population level analyses will reduce the small number statistics problem we currently have when attempting to constrain dark matter using only one stream at a time.

\section{Acknowledgments}
The authors thank Kathryn Johnston, Jake Nibauer, Nathaniel Starkman, and the larger CATS Collaboration for useful discussions.

\software{\texttt{numpy} \citep{numpy:20}, \texttt{Astropy} \citep{Astropy:13, Astropy:18, Astropy:22}, \texttt{jax} \citep{jax:18}, \texttt{numpyro} \citep{Bingham:19, Phan:19}, \texttt{galstreams} \citep{Mateu:2023}}

\bibliography{main}

\appendix

\section{Notation} \label{app:notation}

Here we summarize notation used in equations and probabilistic expressions throughout the article.

\begin{table}[h]
\centering
\begin{tabular}{r l l} 
    Symbol & Domain & Description \\
    \hline \hline
    $\alpha$ & $[0, 1]$ & Mixture weight \\
    $m$ & $[-\infty, \infty]$ & Mean (used instead of $\mu$, which here is proper motion)\\
    $\sigma$ & $(0, \infty]$ & Standard deviation (i.e. square-root of variance) \\
    $\bs{y}$ & $[-\infty, \infty]$ & A vector of data values \\
    \hline
    $\tilde{x}, \tilde{y}, ...$ & --- & Represents the ``true'' (i.e. independent of noise) value of a variable (e.g., $x, y, ...$) \\
    $K, M, N, ...$ & --- & Capitalized Roman characters represent integer numbers \\
    $\{x\}_K, \{y\}_M, ...$ & --- & A collection of values with size specified by the subscript \\
    \spline{K}{x} & --- & A cubic spline function with $K$ knots as a function of the coordinate $x$ \\
    $w_k$ & --- & Spline knot values for the $K$ knots of a spline function \\
    \hline
    $\uniform{x}{a,b}$ & --- & Uniform distribution for $x$ with lower bound $a$ and upper bound $b$ \\
    $\mathcal{N}(x \given m, \sigma^2)$ & --- & Normal distribution for $x$ with mean $m$ and standard deviation $\sigma$\\
    $\truncnorm{x}{m}{\sigma^2}{a,b}$ & --- & Truncated normal distribution for $x$ with mean $m$, standard deviation $\sigma$, and bounds $[a,b]$ \\

\end{tabular}
\caption{}
\label{table:notation}
\end{table}

\section{Application to the GD-1 Stream: Model Choices and Priors}
\label{app:model_math}

In this appendix, we describe our choices for the distributions and parameter priors the
components of our density model.
As described in \secref{full}, the total \pdf used in our density model is a three
component (background, stream, and off-track) mixture given by
\begin{equation}
    p(\true{\bs{y}} \given \bs{\theta}) =
        \alpha_s \, p_s(\true{\bs{y}} \given \bs{\theta}_s) +
        \alpha_b \, p_b(\true{\bs{y}} \given \bs{\theta}_b) +
        \alpha_o \, p_o(\true{\bs{y}} \given \bs{\theta}_o)
\end{equation}
where $\bs{\alpha} = (\alpha_s, \alpha_b, \alpha_o)$ are the mixture weights and
$p_{\{s,b,o\}}(\true{\bs{y}} \given \bs{\theta}_{\{s,b,o\}})$ are the mixture component
distributions.
For our application to GD-1, we choose to model all observed phase-space coordinates
except distance (parallax), so the true phase-space vector is $\true{\bs{y}} =
(\true{\phi}_1, \true{\phi}_2, \true{\mu}_1, \true{\mu}_2, \true{v}_r)$.
For brevity in the equations below, we may drop subscripts from parameters of \pdfs
where it is clear what phase-space coordinate they correspond to --- for example, in the
expression $p(y \given \theta) = \mathcal{N}(y \given m, \sigma^2)$, we should
technically make the generic mean and variance parameters $(m, \sigma)$ have subscripts
$y$ to identify them as related to the coordinate $y$, but we do not do this where it is
clear.
The density distributions for $\true{\phi}_1$, $\true{\phi}_2$, $\true{\mu}_1$, $\true{\mu}_2$, and $\true{v}_r$ are truncated to the domains $\true{\phi}_1 \in [-100, 20]\degree$, $\true{\phi}_2 \in [7.97, 3.35]\degree$, $\true{\mu}_1 \in [-15.10, -4.64]\masyr$, $\true{\mu}_2 \in [-5.42, 1.93]\masyr$, and  $\true{v}_r \in [-500,500] \kms.$

\subsection{Stream} \label{app:stream_math}
As introduced in \secref{stream_choices}, our stream component density model factorizes into a PDF over stream longitude
$\true{\phi_1}$ multiplied by the other
\pdfs, conditioned on stream longitude:
\begin{equation} \label{eqn:stream_pdf}
    p_s(\true{\bs{y}} \given \bs{\theta}_s) =
    p_s(\true{\phi}_1 \given \bs{\theta}_s)
    \, p_s(\true{\phi}_2 \given \bs{\theta}_s)
    \, p_s(\true{\mu}_1 \given \true{\phi}_1, \bs{\theta}_s)
    \, p_s(\true{\mu}_2 \given \true{\phi}_1, \bs{\theta}_s)
    \, p_s(\true{v}_r \given \true{\phi}_1, \bs{\theta}_s) \quad .
\end{equation}

We model the $\phi_1$ density with a mixture of truncated normals with fixed mean locations (spaced by $5\degree$ along $\phi_1$ from $-100\degree$ to $20\degree$),
\begin{equation}
    p_s(\true{\phi}_1 \given \bs{\theta}_s) =
        \sum_i^{25} \alpha_i \, \mathcal{N}(\true{\phi}_1 \given m_i, \sigma^2_i)
\end{equation}
where $m_i = (-100,-95,-90,\ldots,10,15,20)\degree$ are the fixed means of the normals, $\sigma_i$ are the scales of the normal components with truncated normal priors $\truncnorm{\sigma_i}{2.5\degree}{(2.5\degree)^2}{0.5\degree, \infty}$, and $\alpha_i$ are the mixture weights with a uniform Dirichlet prior.

The parameters of the latter four conditional distributions in Equation~\ref{eqn:stream_pdf} are controlled by cubic spline functions, and the remaining parameters of the stream model component are the values of these spline functions at (fixed) knot locations.
We fix the knot locations for all spline functions of $\phi_1$ every $10\degree$ from $-100\degree$ to $20\degree$, inclusive.
We summarize these coordinate distributions and parameters in
Table~\ref{table:gd1_choices}.

The $\phi_2$, $\mu_1$, $\mu_2$, and $v_r$ density distributions are each given by a truncated normal distribution with mean ($m$) and scale ($\sigma$) parameters ($m$ and $\sigma$, respectively) set by spline functions with fixed knot locations,
\begin{equation}
\begin{split}
    p_s(\true{\phi}_2 \given \phi_1, \bs{\theta}_s) &=
        \truncnorm{\true{\phi}_2}{m(\phi_1)}{\sigma^2(\phi_1)}{-7.97\degree,3.35\degree}\\
    p_s(\true{\mu}_1 \given \phi_1, \bs{\theta}_s) &=
        \truncnorm{\true{\mu}_1}{m(\phi_1)}{\sigma^2(\phi_1)}{-15.10\masyr,-4.64\masyr} \\
    p_s(\true{\mu}_2 \given \phi_1, \bs{\theta}_s) &=
        \truncnorm{\true{\mu}_2}{m(\phi_1)}{\sigma^2(\phi_1)}{-5.42\masyr,1.93\masyr} \\
    p_s(\true{v}_r \given \phi_1, \bs{\theta}_s) &=
        \truncnorm{\true{v}_r}{m(\phi_1)}{\sigma^2(\phi_1)}{-500\kms,500\kms}
\end{split}
\end{equation}
where the mean ($m$) and scale ($\sigma$) parameters are set by spline functions with fixed knot locations
\begin{equation}
\begin{split}
    m(\phi_1) &= \spline{K_m}{\true{\phi}_1} \\
    \sigma(\phi_1) &= \spline{K_\sigma}{\true{\phi}_1}
\end{split}
\end{equation}
The knot values $w_{K_m, m}$ and $w_{K_\sigma, \sigma}$ are free parameters.
Because GD-1 is such a well-studied stream, we have the advantage of being able to establish reasonable priors on the $w_{K_m, m}$ knot values.
In detail, we take cubic splines of the PWB proper motion tracks ($\mathcal{S}_{pwb}^{(K_m)}(\true{\phi}_1)$ and evaluate them at our knot locations $\phi_{1,K}$.
These provide the means of our truncated normal distributions at each knot.
The PWB track for radial velocity is less well-measured so we choose a uniform prior instead for those knots.
Mathematically, we represent this by saying that the knot values for the mean functions $w_{K_m, m}$ have truncated normal priors
\begin{equation}
\begin{split}
    \phi_2: \quad &\truncnorm{w_{K_m, m}}{\mathcal{S}_{\textrm{pwb},\phi_2}^{(K_m)}(\phi_{1,K})}{(1\degree)^2}{-7.97\degree,3.35\degree} \\
    \mu_1: \quad &\truncnorm{w_{K_m, m}}{\mathcal{S}_{\textrm{pwb},\mu_1}^{(K_m)}(\phi_{1,K})}{(2\masyr)^2}{-15.10\masyr,-4.64\masyr} \\
    \mu_2: \quad &\truncnorm{w_{K_m, m}}{\mathcal{S}_{\textrm{pwb},\mu_2}^{(K_m)}(\phi_{1,K})}{(2\masyr)^2}{-5.42\masyr,1.93\masyr} \\
    v_r: \quad & \mathcal{U}(w_{K_m, m} \given -500 \kms, 500 \kms) 
\end{split}
\end{equation}
where we have chosen the scales to be to allow some variation on the literature values but not too much, to aid the optimization process.
The knot values for the scale functions $w_{K_\sigma, \sigma}$ also have truncated normal priors
\begin{equation}
\begin{split}
    \phi_2: \quad &\truncnorm{w_{K_{\sigma}, \sigma}}{0.5\degree}{(0.5\degree)^2}{0.05\degree,\infty} \\
    \mu_1: \quad &\truncnorm{w_{K_{\sigma}, \sigma}}{0.5\masyr}{(0.5\masyr)^2}{2.78\times10^{-3}\masyr,\infty} \\
    \mu_2: \quad &\truncnorm{w_{K_{\sigma}, \sigma}}{0.5\masyr}{(0.5\masyr)^2}{2.78\times10^{-3}\masyr,\infty} \\
    v_r: \quad &\truncnorm{w_{K_{\sigma}, \sigma}}{1\kms}{(5\kms)^2}{0.1\kms,100\kms}
\end{split}
\end{equation}
The $2.78\times10^{-3} \masyr$ number comes from converting $0.1 \kms$ to a proper motion at $8\kpc$, to avoid proper motion distributions with widths that approach 0.
We use $K_m = K_\sigma = 13$ knots for the $m$ and $\sigma$
spline functions in proper motions and $K_m = K_\sigma = 9$ for the radial velocities.
The difference comes from the fact that while all our knots are separated by 10\degree, the radial velocity data for likely stream members has a smaller $\phi_1$ range ($-82.29\degree < \phi_1 < 2.60\degree$).

The stream model component parameter vector $\bs{\theta_s}$ therefore consists of:
\begin{itemize}
    \item the scales and mixture weights of the $\phi_1$ normal mixture,
    \item the knot values for spline functions that control the $\phi_2$, $\mu_1$,$\mu_2$, and $v_r$ means, and scales
\end{itemize}
In total, the stream model has 146 free parameters.

\subsection{Background} \label{app:bkg_math}

As with our stream model, our background component density model -- introduced in \secref{background_choices} --
factorizes into a \pdf over stream longitude $\true{\phi_1}$ multiplied by the other
\pdfs, conditioned on stream longitude:
\begin{equation}
    p_b(\true{\bs{y}} \given \bs{\theta}_b) =
    p_b(\true{\phi}_1 \given \bs{\theta}_b)
    \, p_b(\true{\phi}_2 \given \bs{\theta}_b)
    \, p_b(\true{\mu}_1 \given \true{\phi}_1, \bs{\theta}_b)
    \, p_b(\true{\mu}_2 \given \true{\phi}_1, \bs{\theta}_b)
    \, p_b(\true{v}_r \given \true{\phi}_1, \bs{\theta}_b) \quad .
\end{equation}
The parameters of the latter three conditional distributions are controlled by cubic
spline functions, and most of the parameters of the background model are the values of these spline functions at the (fixed) knot locations.
We fix the knot locations for all spline functions of $\phi_1$ below at $(-100, -60, -20, 20)\degree$.
We opt not to do a cross-validation of the knot spacing for this application (because of the computational cost), but we do test the background model with alternate knot spacings of $20\degree$ and $30\degree$.
The former clearly over-fits the data, while the latter gives nearly identical results to the $40\degree$ separations.
This simple test suggests that the total model should not be very sensitive to the precise knot spacings of the background model.
We summarize these coordinate distributions and parameters in
Table~\ref{table:gd1_choices}.

We model the $\phi_1$ density with a mixture of truncated normals with fixed mean
locations (spaced by $40 \degree$ along $\phi_1$),
\begin{equation}
    p_b(\true{\phi}_1 \given \bs{\theta}_b) =
        \sum_i^4 \alpha_i \, \mathcal{N}(\true{\phi}_1 \given m_i, \sigma^2_i)
\end{equation}
where $m_i = (-100, -60, -20, 20)\degree$ are the (fixed) means of the normals, $\sigma_i$ are the scales of the normal components with truncated normal priors $\truncnorm{\sigma_i}{20\degree}{(20\degree)^2}{4\degree, \infty}$, and $\alpha_i$ are the mixture weights with a uniform Dirichlet prior.

The $\phi_2$ distribution is assumed to be uniform --- from visual inspection (top panel
of \figref{gd1_data}), this is a reasonable assumption for the sky region around GD-1.
The density is therefore given by
\begin{equation}
    p_b(\true{\phi}_2 \given \bs{\theta}_b) = \uniform{\true{\phi}_2}{-7.97\degree, 3.35\degree}
\end{equation}
and has no free parameters.

The $\mu_1$ and $\mu_2$ density distributions are both given by a mixture of two
truncated normal distributions,
\begin{equation}
\begin{split}
    p_b(\true{\mu}_1 \given \phi_1, \bs{\theta}_b) &=
    \sum_i^2 \alpha_{\mu_1, i} \,
        \truncnorm{\true{\mu}_1}{m_{\mu_1, i}}{\sigma^2_{\mu_1, i}}{-15.10\masyr,-4.64\masyr} \\
    p_b(\true{\mu}_2 \given \phi_1, \bs{\theta}_b) &=
    \sum_i^2 \alpha_{\mu_2, i} \,
        \truncnorm{\true{\mu}_2}{m_{\mu_2, i}}{\sigma^2_{\mu_2, i}}{-5.42\masyr,1.93\masyr}
\end{split}
\end{equation}
where the parameters of the mixtures (component means $m_i$,
and component scales $\sigma_i$) are controlled by spline functions.
We assume the mixture weights $\alpha_i$ are constant along $\phi_1$.
For each coordinate $\mu_1, \mu_2$, the parameters are treated as cubic spline functions
of $\true{\phi}_1$,
\begin{equation}
\begin{split}
    m_i(\phi_1) &= \spline{K_m}{\true{\phi}_1} \\
    \sigma_i(\phi_1) &= \spline{K_\sigma}{\true{\phi}_1}
\end{split}
\end{equation}
with fixed knot locations $x_k$ for the $K_m$ knots for the mean functions, and $K_\sigma$ knots for the scale functions.
The values of these 4 spline functions at the knot locations, $w_k$, are free parameters of the background model.
The knot values for the mean functions $w_{K_m, m}$ have a uniform prior
$\uniform{w_{K_m, m}}{-10\masyr, 10\masyr}$ and the knot values for the scales
$w_{K_\sigma, \sigma}$ have a truncated normal prior $\truncnorm{w_{K_\sigma, \sigma}}{4 \masyr}{(3\masyr)^2}{2.78\times10^{-3} \masyr, \infty}$.
We use $K_m = K_\sigma = 4$ knots for both the $\mu_1$ and $\mu_2$ mean and scale spline functions.

The $v_r$ probability density is analogous to the proper motion ones:
\begin{equation}
    p_b(\true{v}_r \given \phi_1, \bs{\theta}_b) =
        \sum_i^2 \alpha_{v_r, i} \,\truncnorm{\true{v}_r}{m_{v_r, i}(\phi_1)}{\sigma^2_{v_r, i}(\phi_1)}{-500,500}
\end{equation}
where
\begin{equation}
\begin{split}
    m_{v_r, i}(\phi_1) &= \spline{K_m}{\true{\phi}_1} \\
    \sigma_{v_r, i}(\phi_1) &= \spline{K_\sigma}{\true{\phi}_1}
\end{split}
\end{equation}
with knot values $w_{K_m, m}$ and $w_{K_\sigma, \sigma}$ as free parameters.
The knot values for the mean function $w_{K_m, m}$ have a uniform prior
$\uniform{w_{K_m, m}}{-500\kms, 500\kms}$ and the knot values for the scale function
$w_{K_\sigma, \sigma}$ have a truncated normal prior $\truncnorm{w_{K_\sigma,
\sigma}}{100 \kms}{(100\kms)^2}{0.1 \kms, \infty}$.
As with the proper motions, we use $K_m = 4$ and $K_\sigma = 4$ knots for the $m$ and $\sigma$ spline functions, respectively.

The background model component parameter vector $\bs{\theta}_b$ therefore consists of:
\begin{itemize}
    \item the scales and mixture weights of the $\phi_1$ normal mixture,
    \item the knot values for spline functions that control the $\mu_1$,$\mu_2$, and $v_r$ means and scales,
    \item the mixture weights $\alpha_i$ for the $\mu_1$, $\mu_2$, and $v_r$ truncated normal mixtures.
\end{itemize}
In total, the background model has 62 free parameters.

\subsection{Off-track} \label{app:offtrack_math}
Finally, our off-track component density model -- introduced in \secref{offtrack_choices} -- factorizes into a PDF over the joint ($\true{\phi}_1,\true{\phi}_2$) space, multiplied by the PDFs for the kinematic coordinates, still conditioned on stream longitude $\phi_1$:
\begin{equation}
    p_o(\true{\bs{y}} \given \bs{\theta}_o) =
    p_o(\true{\phi}_1, \true{\phi}_2 \given \bs{\theta}_o)
    \, p_o(\true{\mu}_1 \given \phi_1, \bs{\theta}_o)
    \, p_o(\true{\mu}_2 \given \phi_1, \bs{\theta}_o)
    \, p_o(\true{v}_r \given \phi_1, \bs{\theta}_o)
\end{equation}

We model the ($\phi_1,\phi_2$) density with a mixture of truncated 2D normal distributions with fixed mean locations placed on a grid separated by $3\degree$ in $\phi_1$ and $1\degree$ in $\phi_2$,
\begin{equation}
    p_o(\true{\phi}_1, \true{\phi}_2 \given \bs{\theta}_o)) = \;\sum_i^{390} \alpha_i \:\mathcal{N}(\true{\phi}_1, \true{\phi}_2 | m_{i,1}, m_{i,2}, \sigma^2_{i,1}, \sigma^2_{i,2})
\end{equation}
where $m_{i,1}$ and $m_{i,2}$ are the (fixed) $\phi_1$ and $\phi_2$ positions of the means of the normals, respectively, while $\sigma_{i,1}$ and $\sigma_{i,2}$ are the scales of the normal components in $\phi_1$ and $\phi_2$, respectively.
These scales have truncated normal priors $\truncnorm{(\sigma_{i,1},\sigma_{i,2})}{(3\degree,1\degree)}{(3\degree,1\degree)^2}{(0.3\degree,0.1\degree), (\infty,\infty)}$.
Lastly, $\alpha_i$ are the mixture weights with a uniform Dirichlet prior.

The probability density of the proper motions and radial velocities is similar to that of the stream.
For the stream, we created priors based on the PWB tracks in the three velocity dimensions.
For the off-track component, we create them based on the results of an initial ``stream+background'' VI run.
This ensures that the off-track velocities remain close to those of the stream.
Specifically, letting $\mathcal{S}_{\textrm{stream},\mu_1}^{(K_m)}$, $\mathcal{S}_{\textrm{stream},\mu_2}^{(K_m)}$, and $\mathcal{S}_{\textrm{stream},v_r}^{(K_m)}$ denote the splines for the mean stream velocities, the knot values for the mean off-track functions $w_{K_m,m}$ have truncated normal priors
\begin{equation}
\begin{split}
    \mu_1: \quad &\truncnorm{w_{K_m, m}}{\mathcal{S}_{\textrm{stream},\mu_1}^{(K_m)}(\phi_{1,K})}{(0.25\masyr)^2}{\mathcal{S}_{\textrm{stream},\mu_1}^{(K_m)}(\phi_{1,K}) - 1 \masyr, \: \mathcal{S}_{\textrm{stream},\mu_1}^{(K_m)}(\phi_{1,K}) + 1 \masyr} \\
    \mu_2: \quad &\truncnorm{w_{K_m, m}}{\mathcal{S}_{\textrm{stream},\mu_2}^{(K_m)}(\phi_{1,K})}{(0.25\masyr)^2}{\mathcal{S}_{\textrm{stream},\mu_2}^{(K_m)}(\phi_{1,K}) - 1 \masyr, \: \mathcal{S}_{\textrm{stream},\mu_2}^{(K_m)}(\phi_{1,K}) + 1\masyr} \\
    v_r: \quad &\truncnorm{w_{K_m, m}}{\mathcal{S}_{\textrm{stream},v_r}^{(K_m)}(\phi_{1,K})}{(5\kms)^2}{\mathcal{S}_{\textrm{stream},v_r}^{(K_m)}(\phi_{1,K}) - 20\kms, \: \mathcal{S}_{\textrm{stream},v_r}^{(K_m)}(\phi_{1,K}) + 20 \kms}
\end{split}
\end{equation}

The off-track velocity scales are also drawn from truncated normals centered on the stream velocity scales with standard deviation equal to 1/3 the stream velocity scales.
We truncate the distributions at $0.00278 \masyr$ and $0.1 \kms$ on the lower end for the proper motions and radial velocities, respectively, and twice the stream velocity scale on the upper end.
These are tight priors because we want to ensure that off-track features are capturing local overdensities related to the stream.

The off-track model therefore consists of:
\begin{itemize}
    \item the scales and mixture weights of the $(\phi_1,\phi_2)$ normal mixture,
    \item the knot values for spline functions that control the $\mu_1$,$\mu_2$, and $v_r$ means and scales,
\end{itemize}
In total, the off-track model has 850 free parameters.

\section{Background and Stream Optimization Initialization} \label{app:initialization}

Given the large number of parameters in our GD-1 model, we want to improve computation time by using an initialization that is reasonably close to the true solution.
To do so, we run VI optimization on the stream and background components individually (using data subsets described in \secref{background_choices}) and use the results to initialize the full model.
Here, we provide the initialization choices for these individual runs.

\noindent \textbf{Background}: The background density weights are set to be equal at each knot, with the scales set to $40\degree$.
The weights of each of the two components in the three velocity dimensions are set to 0.5.
We initialize both $\mu_1$ normals with mean $0 \masyr$ and scale $5 \masyr$.
We initialize both $\mu_2$ normals with mean $-3 \masyr$ and scale $3 \masyr$.
We initialize both $v_r$ normals with mean $-100 \kms$ and scale $50 \kms$.

\noindent \textbf{Stream}: The stream density weights ($\bs{\alpha}$ in \appref{stream_math}) are set to be equal at each knot with scales set to 10\degree.
We initialize the $\phi_2,\:\mu_1,\:\mu_2$, and $v_r$ normal means at each node based on the tracks in PWB18. The scales are set to $0.5\degree, \: 0.35 \masyr, \: 0.35 \masyr$, and $4 \kms$, respectively, where the proper motion scales are based on the average proper motion uncertainty of GD-1 members.

\begin{figure*}[th!]
    \centering
    \includegraphics[width=\linewidth]{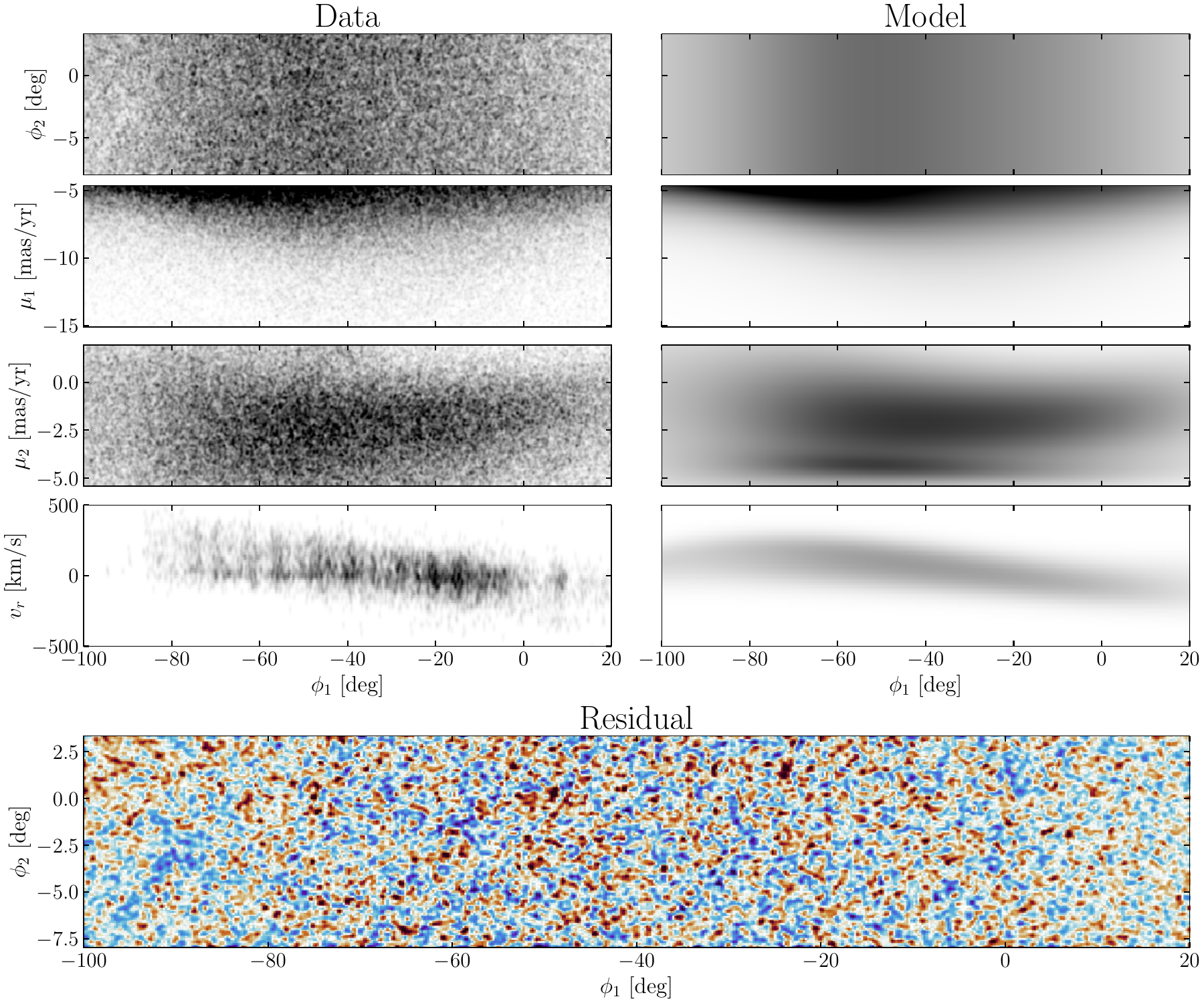}
\caption{The same as \figref{full_model} but for the background model described in \secref{background_choices}.}
\label{fig:background_model}
\end{figure*}

\begin{figure*}[th!]
    \centering
    \includegraphics[width=\linewidth]{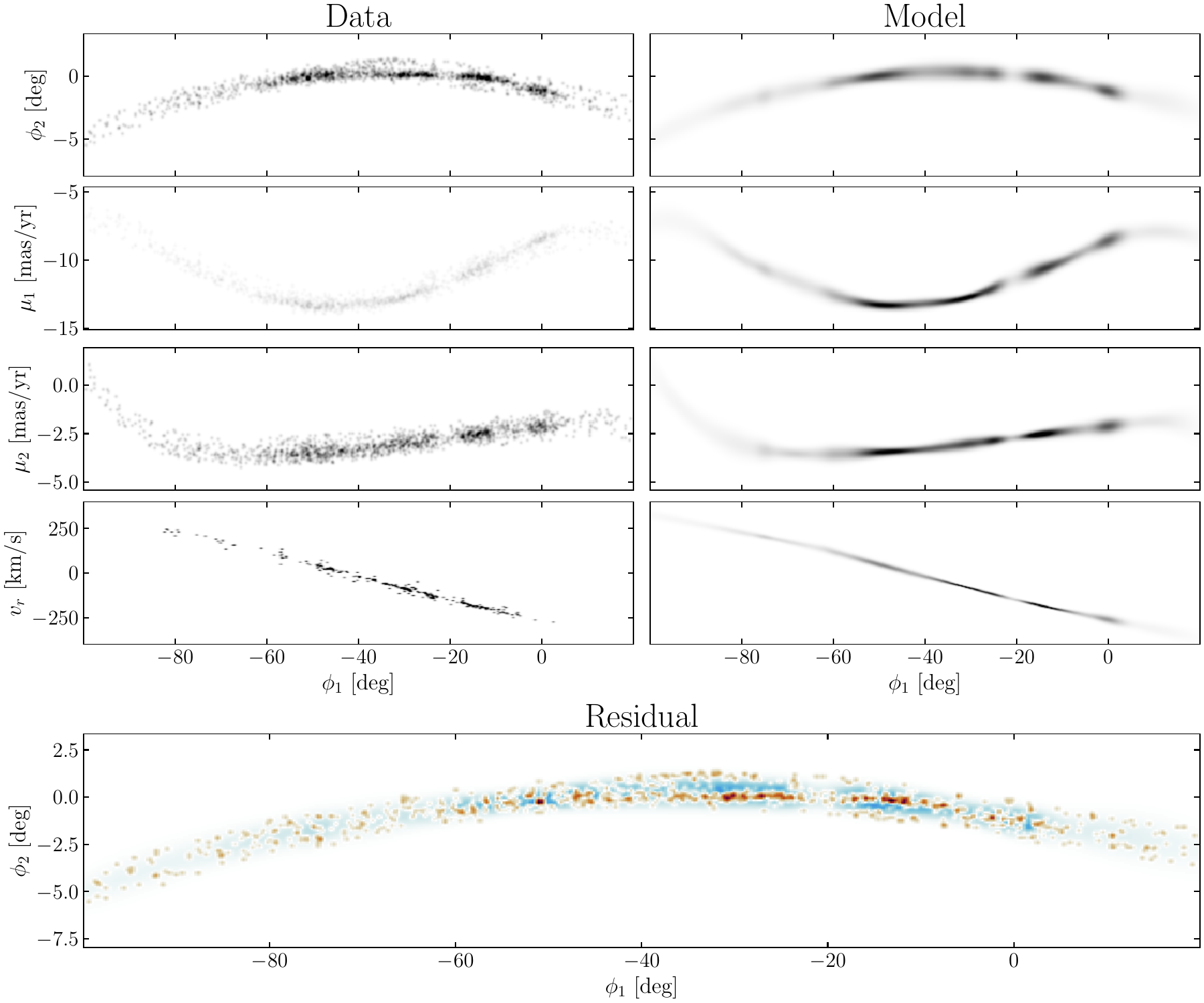}
\caption{The same as \figref{full_model} but for the stream model described in \secref{stream_choices}.}
\label{fig:stream_model}
\end{figure*}

The results of the individual component runs are shown in Figures~\ref{fig:background_model} and \ref{fig:stream_model}.

\end{document}